%% file: Notes.tex
\documentclass[a4paper, english]{article}

\usepackage{silence}
\usepackage{"./technical/UserDefinitions"}
\usepackage{standalone} 
\usepackage{algorithm2e}
\usepackage{nicefrac}
\usepackage{xparse}
\usepackage[normalem]{ulem}
\usepackage{bm}
\usepackage{csquotes}

\graphicspath{{images/}}

\ExplSyntaxOn
\DeclareExpandableDocumentCommand \round { O{0} m }
 { \fp_eval:n { round(#2,#1) } }
\ExplSyntaxOff

\renewcommand*{\algorithmcfname}{}
\makeatletter
\newcommand{\RemoveAlgoNumber}{\renewcommand{\fnum@algocf}{\AlCapSty{\AlCapFnt\algorithmcfname}}}
\newcommand{\RevertAlgoNumber}{\algocf@resetfnum}
\makeatother
\SetAlgoCaptionSeparator{}

\begin{document}
	\begin{titlepage}
	\begin{center}
	
	\vspace*{3 cm}
	{\Large\bfseries Erratic Extremism causes Dynamic Consensus \\}
	{\large\bfseries (a new model for one-dimensional opinion dynamics) \\}

	\vspace{5cm}
	{\Large Dmitry Rabinovich and Alfred M. Bruckstein\\}
	\vspace{1.5cm}

	\capitalize{Center for Intelligent Systems (CIS)}\\[5pt]
	\capitalize{Multi-Agent Robotic Systems (MARS) Laboratory}\\[5pt]
	\capitalize{Computer Science Dept.}\\[5pt]
	\capitalize{Technion Israel Institute of Technology}\\[5pt]
	{3200003, Haifa,
	Israel}\\
	
	\vfill

	\DTMsetup{useregional}
	\DTMtryregional{en}{US}
	\today

	\end{center}
	\end{titlepage}

	\null\vfill
	\begin{abstract}
		A society of agents, with ideological positions, or \enquote{opinions} measured by real values ranging from $-\infty$ (the \enquote{far left}) to $+\infty$ (the \enquote{far right}), is considered.
		At fixed (unit) time intervals agents repeatedly reconsider and change their opinions if and only if they find themselves at the extremes of the range of ideological positions
		held by members of the society. Extremist agents are erratic: they become either more radical, and move away from the positions of other agents, with probability $\varepsilon$, or more moderate, 
		and move towards the positions held by peers, with probability $(1 - \varepsilon)$. The change in the opinion of the extremists is one unit on the real line. We prove that
		the agent positions cluster in time, with all non-extremist agents located within a unit interval. However, the consensus opinion is dynamic. Due to the 
		extremists' erratic behavior the clustered opinion set performs a \enquote{sluggish} random walk on the entire range of possible ideological positions (the real line). 		
		The inertia of the group, the reluctance of the society's agents to change their consensus opinion, increases with the size of the group. 
		The extremists perform biased random walk excursions to the right and left and, in time, their actions succeed to move the society of agents in random directions. The \enquote{far left} agent effectively 
		pushes the group consensus toward the right, while the \enquote{far right} agent counter-balances the push and causes the consensus to move toward the left.

		We believe that this model, and some of its variations, has the potential to explain the real world swings in societal ideologies that we see around us.
	\end{abstract}
	\vfill

	\newpage
	
	\section{Introduction}
		Over the years, social psychologists proposed numerous explanations for the complex behavior emerging in large groups of supposedly intelligent agents, like tribes and nations. They proposed models 
		and principles of individual behavior and some of these models were even amenable to mathematical analysis enabling predictions about long-term behavior and the inevitable emergence of
		surprising global economic or political phenomena.
		
		The ideas of balance theory \mycite{cartwright1956structural} and social dissonance \mycite{festinger1962theory} led to the consideration of several basic mathematical models, attempting to incorporate 
		the idea that individuals, or agents attempt to reach an equilibrium between their drives, opinions and \enquote{local comfort} and those in their neighborhood. They do so by adjusting their position
		(ideological, political, economic, or spatial) to be similar, or comfortably near the position of their neighbors.
		
		\vspace*{\baselineskip}
		Simplified mathematical models for multi-agent interaction consider a group, colony, society or swarm of agents, each agent associated with a quantity which can be a real number, or a vector,
		describing the \enquote{state}, opinion or position of the agent. The state of the whole group (at time $t$) is specified by the vector $\mathbb{X}(t) \triangleq [x_1(t), x_2(t), \ldots, x_N(t)]^T$, 
		where $x_k(t)$ is the state of agent $k$ at time $t$, and the group comprises $N$ agents.
		
		Then, models postulate that, from some initialization $\mathbb{X}(0)$ at time $t=0$, the state of the system evolves, and, if we consider that changes happen at equal intervals
		(arbitrarily set to one), we obtain general discrete time evolution models of the following form
		\begin{equation*}
			\begin{cases}
				\mathbb{X}(t+1) = \Psi(\mathbb{X}(t)) \\
				\mathbb{X}(0) - \text{initial condition}
			\end{cases}
		\end{equation*}
		Here $\Psi$ describes the way each agent $k$, determines its state at time $(t+1)$ given the states of all agents at time $t$.
		
		The inter-agent interaction function $\Psi$ is designed to reflect the assumed influence of agents on their peers. \mycite{degroot1974reaching} postulated that $\Psi$ should be a fixed matrix $A$ 
		acting on $\mathbb{X}$ with columns displaying the influence each agent has on every other agent. Rows of the matrix then display how the next state of agent $k$ at time $(t+1)$ will be computed 
		as a weighted combination of the states of all agents at time $t$. If $A$ is constant (and independent of the state at all times) the vector $\mathbb{X}$ has a linear evolution, with dynamics 
		completely determined only by the eigen-structure of $A$ and the initial state.
		
		When positive entries and convex combination of states are postulated, $A$ is a stochastic matrix, and then one readily has, under quite general conditions, that the system asymptotically achieves
		consensus, i.e. as $t \to \infty$ we have that all $x_k(t)$'s will evolve to have the same value.
		
		\vspace*{\baselineskip}
		This model is highly appealing, however it assumes that each agent always adjusts its state according to a fixed convex combination of its own state and all other states.		
		Since real individuals in any group are well known to posses a certain reluctance in considering far-away positions of others, and tend to stick to their initial opinions, models that took such tendencies 
		into consideration soon emerged. The very popular Hegselmann-Krause (HK) model \mycite{hegselmann2002opinion} postulates that
		\begin{equation*}
			x_k(t+1) = \frac{1}{\abs{\mathcal{N}_k}}\sum\limits_{l \in \mathcal{N}_k} x_l(t),
		\end{equation*}
		where the index set $\mathcal{N}_k \triangleq \{l \enspace : \enspace \norm{x_k(t) - x_l(t)} < \varepsilon_k \}$, i.e. $\mathcal{N}_k$ defines an $\varepsilon_k$-neighborhood of the $k$-th 
		agent position, $x_k(t)$, at time $t$, and $\abs{\mathcal{N}_k}$ is the size of the set $\mathcal{N}_k$.
		
		This model leads, in general, to clusters of agents in local consensus at different state values/positions, a phenomenon often observed in society. Several variations based on this model were put forth in the
		literature and a lot of research is still devoted to study their convergence and properties.
		
		Another interesting variation of the DeGroot model was proposed in \mycite{friedkin1990social}. This model assumes that each agent $k$ remains faithful to its initial position to a certain degree $g_k$, 
		$0 \leq g_k \leq 1$ and	has a susceptibility of $1-g_k$ to be socially influenced by the other agents. The classical linear model then becomes, in a matrix notation:
		\begin{equation*}
			\mathbb{X}(t+1) = G\mathbb{X}(0) + (I - G) A\mathbb{X}(t),
		\end{equation*}
		Here $G$ is a diagonal matrix with $g_k$-s on the main diagonal, and $I$ is the identity matrix. This model leads to a spread of steady state positions that can be predicted by a simple matrix inversion.
		
		Following the footsteps of \mycite{degroot1974reaching}, \mycite{friedkin1990social} and \mycite{hegselmann2002opinion}, a considerable number of interesting \enquote{opinion dynamics}, 
		\enquote{multi-agent} and \enquote{consensus}/\enquote{gathering} models have been proposed. 
		Over the years the research in the field split into several branches. Today researchers of Autonomous Swarms and Swarm Intelligence invent local interaction models to achieve 
		\enquote{gathering}, \enquote{geometric consensus}, \enquote{collective area sweeps} and \enquote{cooperative search and pursuit}  with simple autonomous mobile agents (see e.g. \mycite{reynolds1987flocks}, 
		\mycite{dudek1993taxonomy}, \mycite{vicsek1995novel}, \mycite{ando1999distributed}, \mycite{citeulike:606465}, \mycite{jadbabaie2003coordination},	\mycite{olfati2004consensus}, \mycite{csahin2004swarm}). 
		An overview of this field is provided in \mycite{barel2016}. Computer Scientists are interested in agreement and common
		knowledge in distributed computer networks (\mycite{halpern1990knowledge}, \mycite{shoham1995social}, \mycite{flocchini2012distributed}), 
		while communication engineers consider distributed coordination and collaboration in large, ad-hoc networks of \enquote{cellular-phone} agents (\mycite{krishna1997cluster}, \mycite{chen2002span}, \mycite{chong2003sensor}). 
		Biologists analyze and try to understand and model colonies of ants, flocks of starlings, schools of fish and swarms of locusts \mycite{citeulike:6170389}, \mycite{citeulike:606465}, 
		\mycite{couzin2003self}, \mycite{sumpter2006principles}.
		
		Social science researchers continue to be interested in simulating and analyzing human agent interactions, voting patterns and social opinion dynamics. A recent survey by \mycite{lorenz2017modeling}
		nicely presents the advancements and clearly describes some of the issues of interest in the field. Stochastic models, explicitly dealing with random behavior of agents with parameters
		probabilistically characterizing their open-mindedness (the agent's probability of changing/reconsidering opinions), are currently being investigated. In his concluding remarks, Lorentz states
		\clearpage
		\begin{quote}
			``Agent-based models for the evolution of ideological landscapes are still in infancy and it remains to show if they can add interesting insight to political dynamics."
		\end{quote}
		\rightline{{\mycite{lorenz2017modeling}, page 265}}
		\vspace {4mm}
		
		In this vein, we here propose a new, probabilistic, opinion dynamics model, in part based on some early ideas of \mycite{festinger1954theory}. He introduced a qualitative social psychology theory, 
		supported by a vast corpus of data collected. The theory suggests that the majority of agents hold neutral opinion on subjects at hand. This majority is rather unmoved by extreme opinions, while
		the \enquote{extremists} are unstable and tend to fluctuate, moving most probably in the direction of a social norm.
		
		We model opinions or ideological positions as real numbers and allow only extreme agents to change opinions at discrete times by a constant quantum value arbitrarily set to one in any direction. 
		Changes in the positions of the \enquote{extremists} in the direction of the \enquote{social norm}, (represented by all agents except the two \enquote{extremists}), are assumed to be highly probable. 
		In the opposite direction the erratic \enquote{extremists} may move, but with smaller probabilities. We show that for any initial spread of agent opinions, a consensus opinion arises. 
		The \enquote{core} group in consensus spreads over an interval of size smaller than the quantum change in the opinion of the extreme agents. The \enquote{core} is not stationary and, over time, 
		moves at random. In the society of agents \enquote{extremist} is not a sticky label. From time to time one of the \enquote{extremists} becomes a part of the \enquote{core} of normal agents; 
		a previously \enquote{normative} moderate agent finds itself to be at one of the extremes. It is these role-changes between \enquote{extremists} and \enquote{moderates} that moves the \enquote{core}
		over time.
	
		This paper is organized as follows. \autoref{Section_Model} presents the mathematical model of opinion dynamics and states our main results. 
		\autoref{Section_Basic_Facts} reviews and proves some basic facts about biased random walks. \autoref{Section_Analyzing_Gathering_Process} analyzes the 
		gathering process by first considering a unilateral case in which we assume that only one extremal agent is active, then a decoupling trick enables us to use the unilateral results for
		the analysis of the problem when both extremal agents are in action. \autoref{Section_Simulations} presents extensive simulation results confirming the theoretical predictions and showing
		that our bounds are quite loose due to the need to decouple the action of the extremal agents in order to enable the theoretical results. The final \autoref{Section_Conclusions} discusses possible
		interesting extensions of the model presented along with some initial simulation results in two dimensions.
	
	\newpage
	\RemoveAlgoNumber

	\section{Model Description}\label{Section_Model}
		Suppose a set of point agents, the individuals in the society, called $p_1, p_2, \ldots, p_N$ are, at the beginning of time, i.e. at $t=0$, on the real line (the range of positions or opinions) at locations
		$x_1(0), x_2(0), \ldots, x_N(0) \in \mathbb{R}$. The agents are identical and indistinguishable points and perform the following algorithm :

		\RestyleAlgo{boxruled}
		\LinesNumbered
		\begin{algorithm}[ht]
		  \caption*{Agent decision rule\label{alg} ($\varepsilon \in [0, \frac{1}{2})$):}
		  \DontPrintSemicolon
		  For $p_k$ located at $x_k(t)$ at discrete time $t$ define intervals $pR \triangleq (x_k(t), \infty)$ and $pL \triangleq (-\infty, x_k(t))$. \\
		  \uIf{in both intervals $pL$ and $pR$ there are other agents}
		  {
			$x_k(t + 1) = x_k(t)$, i.e. $p_k$ stays put.
		  }
		  \Else
		  {
			\uIf{ $pR$ is empty }
			{
				$p_k$ makes a probabilistic jump, setting \\
				$x_k(t+1) = 
					\begin{cases}
						x_k(t) + 1 & , \text{w.p. } \varepsilon \\
						x_k(t) - 1 & , \text{w.p. } (1-\varepsilon)
					\end{cases}$
			}
			\uIf{ $pL$ is empty }
			{
				$p_k$ makes a probabilistic jump, setting \\
				$x_k(t+1) = 
					\begin{cases}
						x_k(t) + 1 & , \text{w.p. } (1-\varepsilon) \\
						x_k(t) - 1 & , \text{w.p. } \varepsilon
					\end{cases}$
			}
		  }
		\end{algorithm}
		
\begin{figure}[h!]
			\centering
			\input{images/Jumps}
			\caption{N agents on the line}
			\label{fig:1}
		\end{figure}
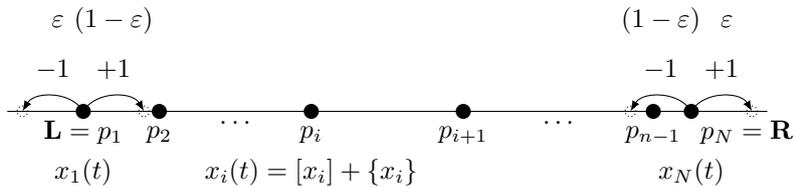

		Under the rule defined above only the two agents with extremal positions $x_{min}(t)$ and $x_{max}(t)$ will move, and their tendency will be to approach the agents in between.
		After each jump, carried out at discrete integer times, we rename the identical agents to have them always indexed in the increasing order of their $x$-locations. Hence at all 
		discrete time instances	$t=1,2, \ldots$ we have the ordered agents $\{p_1, p_2, \ldots, p_N\}$ with $x_1(t) \leq x_2(t) \leq \ldots \leq x_N(t)$, where $p_1$ and $p_N$ are
		extremists and	probabilistic jumps will be carried out by extremists only (see \autoref{fig:1}).
		
		The process defined above evolves the constellation of points in time and we clearly expect that a gathering of the agents will occur, since extremal agents are
		probabilistically \enquote{attracted} toward their peers. 
		
		Indeed, if $\varepsilon$ would be exactly zero, the deterministic jumps carried out by the extremal agents $p_1 \equiv p_L$ and $p_N \equiv p_R$
		would always be toward the interior of the interval $(x_1, x_N)$, shortening it while $(x_N(t)-x_1(t)) > 1$.		
		However, note that when the $[x_1(t), x_N(t)]$ interval reaches a value of $1$ or less, interesting things start to happen, since $p_1$ and $p_N$ while jumping cross each other,
		in such a way that the spread about the (\textbf{time invariant}, in this case) centroid of points may increase and decrease in a way that depends on the specific spread of
		the initial point locations' \textbf{fractional parts}. We therefore expect similar things to happen when randomness is introduced as $\varepsilon$ increases from 0 towards $\nicefrac{1}{2}$.
		For the time being, for simplicity, shall we assume that the fractional parts of the distinct initial locations $x_1(0), x_2(0), \ldots, x_N(0)$ are all different.
		
		\bigskip
		\noindent
			If $\varepsilon = 0$ we have the constellation at time $t$, $\{p_1, p_2, \ldots, p_N\}_t$ described by the ordered set of point locations $x_1(t) < x_2(t) < \ldots < x_N(t)$ and
			their centroid and variance behave as follows: for the centroid arbitrarily chosen to be $0$ at time $0$ we have
			\begin{equation*}
				C(t+1) \triangleq \frac{1}{N} \sum\limits_{i=1}^N x_i(t+1) = \frac{1}{N} \sum\limits_{i=2}^{N-1}x_i(t) + \frac{1}{N}\big(x_1(t) + 1 + x_N(t)-1\big) = C(t), 
			\end{equation*}
			and $C(t+1) = C(t) = \ldots C(0) \triangleq 0$, hence the centroid is an evolution invariant.
			The variance of the constellation about the centroid at $0$ is $\sigma^2(t) = \frac{1}{N} \sum\limits_{i=1}^N x_i^2(t)$, therefore
			\begin{equation*}
				\begin{array}{r c l}
					\sigma^2(t+1) & = & \frac{1}{N} \sum\limits_{i=2}^{N-1}x_i^2(t) + \frac{1}{N}\big(x_1^2(t) + 2x_1(t) + 1 + x_N^2(t) - 2x_N(t) + 1\big) \\
					& = & \frac{1}{N} \sum\limits_{i=1}^N x_i^2(t) - 2\frac{1}{N}\big[(x_N(t)-x_1(t))-1\big] \\
					& = & \sigma^2(t) - \frac{2}{N}\big[(x_N(t)-x_1(t))-1\big]
				\end{array}
			\end{equation*}
			While $(x_N - x_1) > 1$ the variance monotonically decreases, however when $(x_N - x_1) \leq 1$ we have $\sigma^2(t+1) > \sigma^2(t)$. Hence after gathering, or reaching
			consensus (i.e. when $\abs{x_N - x_1} \leq 1$), oscillations in $\sigma^2(t)$ subsequently occur, but the constellation remains gathered around 0.

		\bigskip
		\noindent
		For the probabilistic case we expect a somewhat similar behavior. We shall see that a \enquote{dynamic} consensus is reached.
		Agents on a line behaving according to the probabilistic rule discussed above evolve to a dynamic constellation that is \enquote{gathered}
		and the group of agents move on the line as follows:
		\begin{enumerate}[label={\arabic*)}]
			\item
				For a given $\varepsilon$, $0 < \varepsilon < \nicefrac{1}{2}$, we have
				\begin{equation*}
					C(t+1) = C(t) + 
					\begin{cases}
						\frac{2}{N},  & \text{with probability } \varepsilon(1 - \varepsilon) \\
						0, & \text{with probability } 1-2\varepsilon(1 - \varepsilon) \\
						-\frac{2}{N},  & \text{with probability } \varepsilon(1 - \varepsilon)
					\end{cases}
				\end{equation*}
			\item
				The \enquote{core} group of moderate agents, i.e. $\{p_2, p_3, \ldots, p_{N-1}\}$ eventually gathers to reside within a \enquote{dynamic} interval of length less than one.
			\item
				The extremal agents $p_1$ and $p_{N}$ perform random excursions to the left and right of the core group, with motion biased towards the core. Their bias ensures that they will be mostly near the core, 
				the total distance between them being a sum of random variables, one always less than 1 and two others bounded by positive random variables with a geometric distribution.
		\end{enumerate}
	
	\newpage	
	\section{Some Basic Facts About Random Walk} \label{Section_Basic_Facts}
		In order to analyze the gathering process due to the random behavior of the extremal points ($p_L \triangleq p_1$ and $p_R \triangleq p_N$) in case $\varepsilon > 0$ we need to first recall some 
		basic facts about random walks on the line.
		Suppose an agent performs a (biased) random walk from an initial location (denoted by $x(0) = 0$) on the real line, making, at discrete time instants $t=0, 1, 2, \ldots$ moves
		to the left with probability $(1-\varepsilon)$ and to the right with probability $\varepsilon$. If $\varepsilon = \nicefrac{1}{2}$ the walk is the unbiased, symmetric random walk, 
		while $\varepsilon < \nicefrac{1}{2}$ biases the motion of the agent towards the left. Let us define $\alpha$ as the positive departure of $(\varepsilon)$ and $(1 - \varepsilon)$ from $\nicefrac{1}{2}$, i.e.
		\begin{equation*}
			\varepsilon = \frac{1}{2} - \alpha \Leftrightarrow 1-\varepsilon = \frac{1}{2} + \alpha.
		\end{equation*}
		Clearly, 
		$\alpha \in (0, \nicefrac{1}{2})$, since we assume $0 < \varepsilon < \nicefrac{1}{2}$. In this notation $\alpha$ quantifies the bias towards left of the agents' motion and we have the following results.
		
		\subsection{\textbf{The probability of reaching (-1) from 0.}}
			The probability that the agent hits $(-1)$ is given by the following expression:
			\begin{equation*}
				\begin{array}{r c l}
					\probability{\text{walk hits $(-1)$}} & = & \sum\limits_{k=0}^{\infty} \probability{\text{walk hits $(-1)$ at $(2k + 1)$ for the first time}} \\
					& = & \sum\limits_{k=0}^{\infty} P\left(\parbox{19em}{step to the left after making $k$ steps to the right and $k$ steps to the left in any order, 
						i.e. returning to $0$, without having been at $(-1)$}\right) \\
					& = & \sum\limits_{k=0}^{\infty} \left(\frac{1}{2} + \alpha\right)\cdot C_k \left(\frac{1}{2} - \alpha\right)^k\left(\frac{1}{2} + \alpha\right)^k
				\end{array}
			\end{equation*}
			Here $C_k$ counts the number of possible paths of length $k$ from $0$ to $0$ never reaching $(-1)$, which is given by the $k^{th}$ Catalan number.
			
			It is well known \mycite{stanley2015catalan}, \mycite{hilton1991catalan} that, the generating function of the series $\{C_k\}$ is given by :
			\begin{equation}
				\sum\limits_{k=0}^{\infty} C_k x^k = \sum\limits_{k=0}^{\infty} \frac{1}{k+1}\binom{2k}{k} x^k = \frac{1 - \sqrt{1-4x}}{2x}
			\end{equation}
			
			Hence we have, for $\alpha > 0$,
			\begin{equation*}
				\probability{\text{walk hits $(-1)$}}
				= \left(\nicefrac{1}{2} + \alpha\right) \cdot \frac{1 - \sqrt{1-4(\nicefrac{1}{4}-\alpha^2)}}{2(\nicefrac{1}{2}+\alpha)(\nicefrac{1}{2}-\alpha)}
				= \frac{\left(\nicefrac{1}{2} + \alpha\right) (1-2\alpha)}{(\nicefrac{1}{2}+\alpha)(1-2\alpha)}
				= \mathbf{1}
			\end{equation*}
			This is totally expected : a left biased random walk will almost surely (i.e. with probability $1$) reach $(-1)$, when started at $0$. 
			
		\subsection{\textbf{The probability of reaching (+1) from 0.}}
			We have, similarly :
			\begin{equation*}
				\begin{array}{r c l}
					\probability{\text{walk hits $(+1)$}}
					& = & \sum\limits_{k=0}^{\infty} P\left(\parbox{19em}{walk hits $(+1)$ at step  $2k+1$ for the first time}\right) \\
					& = & \sum\limits_{k=0}^{\infty} P\left(\parbox{19em}{last step to the right after making $k$ steps to the left and $k$ steps to the right 
						(i.e. returning to $0$) without having been at $(+1)$}\right) \\
					& = & \sum\limits_{k=0}^{\infty} (\frac{1}{2} - \alpha)\cdot C_k (\frac{1}{2} + \alpha)^k(\frac{1}{2} - \alpha)^k \\
					& = & \big(\frac{1}{2} - \alpha\big) \frac{1-2\alpha}{(\frac{1}{2}-\alpha)(1+2\alpha)} = \frac{1-2\alpha}{1+2\alpha} < \mathbf{1}
				\end{array}
			\end{equation*}
			
			\noindent
			Hence, while the walk almost surely reaches $(-1)$, there is a non-zero probability, given by $1 - \frac{1-2\alpha}{1+2\alpha} = \bm{\frac{1-2\varepsilon}{1-\varepsilon}}$
			of never reaching $(+1)$.
			
		\subsection{\textbf{The expected number of steps to first reach (-1).}}
			Using the generating function for $\{C_k\}$ we can readily calculate the expected number of steps to reach $(-1)$ from $0$. Hence we have the following, (quite well known) result :
			\begin{equation*}
				\begin{array}{r c l}
					\expectation{\text{steps to first hit } (-1)}
					& = & \sum\limits_{k=0}^{\infty} (2k+1) \cdot \probability{\text{walk hits $(-1)$ at step $2k+1$}} \\
					& = & \sum\limits_{k=0}^{\infty} (2k+1)(\frac{1}{2} + \alpha)\cdot C_k (\frac{1}{2} - \alpha)^k(\frac{1}{2} + \alpha)^k \\
					& = & (\frac{1}{2} + \alpha) \sum\limits_{k=0}^{\infty} (2k+1) (\frac{1}{4}-\alpha^2)^k C_k
				\end{array}
			\end{equation*}

			\noindent
			To compute this value explicitly we use
			\begin{equation*}
				\sum\limits_{k=0}^{\infty} k C_k x^{k - 1}
				= \frac{\mathrm{d}}{\mathrm{d}x}\left(\sum\limits_{k=0}^{\infty} C_kx^{k} \right)
				= \frac{\mathrm{d}}{\mathrm{d}x}\bigg(\frac{1 - \sqrt{1-4x}}{2x}\bigg)
				= \frac{1 - 2x - \sqrt{1-4x}}{2x^2\sqrt{1-4x}}
			\end{equation*}
			hence we have,
			\begin{equation}
				\sum\limits_{k=0}^{\infty} k C_kx^{k} = \sum\limits_{k=0}^{\infty} \frac{k}{k + 1} \binom{2k}{k} x^k = \frac{1 - 2x - \sqrt{1-4x}}{2x\sqrt{1-4x}}
			\end{equation}
			
			\noindent
			This yields, setting $x$ to $(\frac{1}{2} + \alpha) (\frac{1}{2} - \alpha)$, after some algebra,
			\begin{equation}
				\expectation{\text{steps to first hit (-1)}} = \frac{1}{2\alpha} = \frac{1}{1 - 2\varepsilon}
			\end{equation}		
			
			Of course, the expected number of steps to reach $(+1)$ is infinite, since there is a strictly positive probability given by $\frac{1-2\varepsilon}{1-\varepsilon}$ of never getting there.			
			But, we know \textit{for sure} that the biased random walk (more likely moving to the left) will reach $(-1)$ from $0$ in the, above calculated, finite expected number of steps.
			
		\subsection{The expected farthest excursion to the right on the way from 0 to first reaching (-1).}\label{Result3}
			Another result that we shall need in analyzing the evaluation of the agents' behavior is the following result on the excursions that biased random walks make in the direction opposite to their 
			preferred direction :
			the expected farthest excursion to the right on the way from $0$ to first reaching ($-1$) in the \textit{left biased} random walk is bounded by
			\begin{equation*}
				\begin{array}{r c l}
					\expectation{\text{farthest right excursion}}
					& \leq & \sum\limits_{k=0}^{\infty} k \cdot P\left(\parbox{11em}{walk makes $k$ right steps and $(k+1)$ left steps to first reach (-1)}\right) \\
					& = & \sum\limits_{k=0}^{\infty} k C_k \left(\frac{1}{2} + \alpha\right)\left(\frac{1}{2} - \alpha\right)^k\left(\frac{1}{2} + \alpha\right)^k \\
				\end{array}
			\end{equation*}
			The above inequality can be explained as follows: any excursion that starts at $0$ and eventually ends in $-1$ is necessarily of the odd length $2k+1$ for some $k$. No matter what the actual order of steps
			is, the walk makes $k$ steps to the right and $k+1$ steps to the left (with obvious limitations on the order of the steps). Therefore, the farthest to the right such an excursion could get is a 
			distance $k$ from $0$. Hence, left-hand side of the inequality above is a clear upper bound.
			
			\noindent
			Using the previously established relation $\sum\limits_{k=0}^{\infty} k C_k x^{k} = \frac{1 - 2x - \sqrt{1-4x}}{2x\sqrt{1-4x}}$	we obtain
			\begin{equation} \label{Eq:FarthestExcursion}
				\begin{array}{r c l}
					\expectation{\text{farthest right excursion}} 
					& \leq & \sum\limits_{k=0}^{\infty} k C_k \left(\frac{1}{2} + \alpha\right)\left(\frac{1}{2} - \alpha\right)^k\left(\frac{1}{2} + \alpha\right)^k \\ [9pt]
					& = & \left(\frac{1}{2} + \alpha\right)\sum\limits_{k=0}^{\infty} k C_k \left(\frac{1}{2} - \alpha\right)^k\left(\frac{1}{2} + \alpha\right)^k \\ [9pt]
					& = & \left(\frac{1}{2} + \alpha\right)\frac{\left(\frac{1}{2} - \alpha\right)^2}{2\alpha\left(\frac{1}{2} -\alpha\right)\left(\frac{1}{2}+\alpha\right)}
						= \frac{\frac{1}{2} - \alpha}{2\alpha}
						= \bm{\frac{\varepsilon}{1-2\varepsilon}}
				\end{array}
			\end{equation}				
	
	\newpage
	\section{Analysis of the Dynamic Gathering Process} \label{Section_Analyzing_Gathering_Process}
		\subsection{Unilateral Action Results}
			In order to analyze the gathering process, let us first consider a one sided version where only the rightmost agent moves at each moment and all other agents stay put. Furthermore assume
			that to the left of $p_1$ at $t=0$ we put a \enquote{beacon agent} $p_0$ at $x_0(0) < x_1(0)$. The rightmost agent at times $t = 1, 2, 3, \ldots$ makes a unit jump to the left with 
			high probability $(1-\varepsilon)$, or a jump to the right with probability $\varepsilon$. Suppose the agents are initially located at:
			$x_1(0), x_2(0), \ldots, x_{N-1}(0), x_N(0)$. Clearly the rightmost agent $p_R \equiv p_N$ will first reach, with probability $1$, $(x_N(0)-1)$ in $\frac{1}{1-2\varepsilon}$
			expected number of steps, then from $(x_N(0)-1)$ it will reach a.s. $(x_N(0)-2)$ in further $\frac{1}{1-2\varepsilon}$ expected steps etc. until, at some point it will jump over
			$x_{N-1}(0)$ to land somewhere in the interval $(x_{N-1}(0) - 1, x_{N-1}(0))$, making the agent at $x_{N-1}(0)$ the rightmost agent. This will happen with probability $1$, after a number
			of steps, which we shall denote as $T_{\text{jump}}$, having the expected value of $\frac{1}{1-2\varepsilon}\big[\lfloor x_N(0) - x_{N-1}(0) \rfloor + 1\big]$ number of steps.
			
			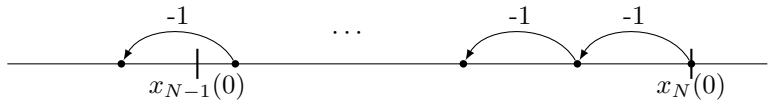
\begin{figure}[h!]
				\centering
				\input{images/JumpSequence}
				\caption{The agent $p_N$ will jump over $p_{N-1}$ after an expected number of steps equal to $\frac{1}{1-2\varepsilon}\big[\lfloor x_N(0) - x_{N-1}(0) \rfloor + 1\big]$}
				\label{fig:2}
			\end{figure}
			
			Now it will be the turn of the former $p_{N-1}$ agent, which is now \enquote{renamed} $p_N \equiv p_R$, to start its biased random walk and it will reach $(x_{N-1}(0)-1)$ in $\frac{1}{1-2\varepsilon}$
			expected steps (clearly jumping over at least the \enquote{current} position of the \enquote{former} moving agent) to land in the interval $(x_1(0)-1, x_N(T_{jump}))$ defined by 
			the \enquote{renamed} agents $(p_1, p_2, \ldots, p_{N-1})$. Clearly the new rightmost agent (which might be the former random walker or another agent located to the left of $x_{N-1}(0)$ in the 
			initial configuration) will do the same.
			
			\textbf{Recall that we assume, for simplicity, that agents' initial locations have all distinct fractional parts, so that one agent will never land on top of another!}
			
			From the above description it is clear that the \enquote{erratic extremist} random walk of rightmost agents will eventually \enquote{sweep} all the agents towards the left, and in a finite expected 
			number of steps equal to :
			\[
				\expectation{T_{\{x_0(0), x_1(0), \ldots, x_N(0)\}}} = \frac{1}{1-2\varepsilon} \sum\limits_{k=1}^{N} (\lfloor x_k(0) - x_0(0) \rfloor + 1),
			\]
			all agents will be to the right of the \enquote{beacon} $p_0$ after having jumped over $x_0(0)$ exactly once, making the \enquote{beacon} $p_0$ the rightmost agent for the first time!
			
			Indeed, note that while jumping one over the other (to the left) all the agents to the right of $p_0$ will have carried out (perhaps with interruptions due to reordering, following jumps over the
			agent called $p_{N-1}$) a biased random walk
			from their initial locations $x_1(0), x_2(0), \ldots, x_N(0)$ until each one of them, for the first time, jumped over the fixed \enquote{beacon} point $p_0$ at $x_0(0)$. Subsequently, the agents will 
			stop and wait for the \enquote{beacon} $p_0$ to become the rightmost agent. This will happen when the last of all the agents (that were $p_0$'s initial right neighbors) completes its random walk by 
			jumping over $p_0$.
			
			An important byproduct of this analysis is the fact that, the moment after the last right neighbor jumps over $p_0$, all the other agents have made \textbf{exactly one left jump} over $p_0$ at $x_0(0)$,
			hence all the agents will be located in the interval $(x_0(0)-1, x_0(0)]$. Therefore we proved:
			
			\begin{theorem}
				If $p_0, p_1, \ldots, p_N$ are located at $t=0$ at $x_0(0), x_1(0), \ldots, x_N(0)$ with $(x_0(0) < x_1(0) < \ldots < x_N(0))$, and the rightmost agent performs random walk biased toward
				the left with probability of a left unit jump of $(1-\varepsilon)$, the agents first gather to the interval $(x_0(0)-1, x_0(0)]$, with probability $\mathbf{1}$, in a finite expected 
				number of steps given by
				$$\expectation{T_{\{x_0(0), x_1(0), \ldots, x_N(0)\}}} = \frac{1}{1-2\varepsilon} \sum\limits_{k=1}^{N} (\lfloor x_k(0) - x_0(0) \rfloor + 1)$$
			\end{theorem}
			
			Note that we could have chosen in this description the \enquote{beacon} to be the leftmost agent $p_1$ located at $x_1(0)$ and then in a finite expected time of
			\begin{equation*}
				\expectation{T_{\{x_1(0), x_2(0), \ldots, x_N(0)\}}} = \frac{1}{1-2\varepsilon} \sum\limits_{k=2}^{N} (\lfloor x_k(0) - x_1(0) \rfloor + 1)
			\end{equation*}
			the agent $p_1$ becomes the rightmost agent. If, beyond the \enquote{first gathering} to the left of $p_1$, the process continues indefinitely, the group of agents will be pushed to the left 
			due to the rightmost agent's actions with an average speed of about $\nicefrac{1-2\varepsilon}{N}$.
			
			Note also that we have the corresponding symmetric result for agent groups where only the leftmost agent is moving and it sweeps all agents, by the action of its biased random walk, towards the right,
			after gathering the group to an interval of length bounded by $1$.

		\subsection{Bilateral Action Results}
			So far we have seen that a unilateral random-walk, biased toward the group of agents, carried out either by the rightmost or by the leftmost agent results in gathering the agents into a
			cluster with a span upper bounded by 1 (i.e. the step size). Something slightly more complex happens when both extremal agents are jointly herding the group. Of course we expect gathering to happen,
			and even faster	than in the case when only one extremal agent is at work. This is indeed the case, however the simultaneous work of the extremal agents leads to interactions that slightly 
			complicate the proofs.
			
			Suppose we have a constellation of agents $p_1, p_2, \ldots, p_N$ located at time $t=0$ at $x_1(0) < x_2(0) < \ldots < x_N(0)$, as before. The \enquote{erratic extremists}, the leftmost
			and rightmost agents $p_L \triangleq p_1$ and $p_R \triangleq p_N$ perform biased steps by simultaneously jumping, towards the agents $\{p_2, p_3, \ldots, p_{N-1}\}$ with probability 
			$(1-\varepsilon)$ or away from them with probability $\varepsilon$. 
			
			The results below represent the main contribution of this paper. \autoref{TheoremGatheringToUnitInterval} states that if the internal agents are gathered in an interval smaller than the step size, 
			they never spread beyond this size. \autoref{thm:FiniteExpectedTimeToHalfShrink} bounds the expected time to shrink the excess distance, beyond one, between $p_2$ and $p_{N-1}$ (the internal agent span) 
			by one half.  
			\autoref{thm:DroppingUnder1} then uses the fact that, once less than 2, the distances $\abs{x_{N-1}(t) - x_2(t)}$ can only take a finite set of values, to show that the inner agents gather to an
			interval of length less than $1$ in finite expected time. \autoref{thm:IntervalDistributions} uses the bounds on the expected excursions of biased random walks in the direction opposite to the bias
			to prove that, with high probability, the total span of all the agents will have a small value as the process continues to evolve after the \enquote{core} gathered.
			
			\begin{theorem}\label{TheoremGatheringToUnitInterval}
				Suppose at $t=T$ the internal agents $\{p_2, p_3, \ldots, p_{N-1}\}$ are all close, so that $x_{N-1}(T) - x_2(T) \leq 1$, then $x_{N-1}(T+1) - x_2(T+1) \leq 1$.
				Hence for all $t > T$ we will have $x_{N-1}(t) - x_2(t) \leq 1$.
			\end{theorem}
			\begin{proof}
				Assume $x_{N-1}(T) - x_2(T) \leq 1$. Designate by $A_L$ and $A_R$ the agents $x_2(T)$ and $x_{N-1}(T)$, respectively. After jumps by extremal agents we can have at
				$t=T+1$ the following cases:
				\begin{itemize}
					\item
						$A_L$ and $A_R$ both remained internal. Then all the internal agents are still inside the interval $[x_2(T), x_{N-1}(T)]$ with assumed length of at most one.
					\item
						$A_L$ and $A_R$ both became extremal. This case is even simpler: all the internal agents at time $T+1$ are now strictly inside the interval $[x_2(T), x_{N-1}(T)]$ with assumed 
						length of at most one. 
					\item
						Either $A_L$ or $A_R$ only became an extremal agent. Assume w.l.o.g. that agent $A_L$ at location $x_2(T)$ became extremal, i.e. $x_1(T+1) = x_2(T)$. In this case all the internal agents
						are contained in either $[x_2(T), x_{N-1}(T)]$ (because the left extremal agents moved into it, see \autoref{TheoremGatheringNearIllustration}) 
						or $\linebreak{[x_2(T), x_{1}(T) + 1]}$ (because the left extremal agent over-jumped all the previous internal agents, see \autoref{TheoremGatheringFarIllustration}).
						In both cases, the interval containing \textit{new} internal agents is of length at most one.
				\end{itemize}
				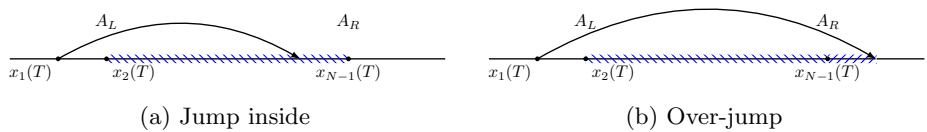
\begin{figure*}[h]
					\centering
					\begin{subfigure}[b]{0.48\textwidth}
						\resizebox{\linewidth}{!}{\input{images/InternalAgentsGatheredNearJump}}
						\caption[]
						{{\small Jump inside}}
						\label{TheoremGatheringNearIllustration}
					\end{subfigure}
					\hfill
					\begin{subfigure}[b]{0.48\textwidth}
						\resizebox{\linewidth}{!}{\input{images/InternalAgentsGatheredFarJump}}
						\caption[]
						{{\small Over-jump}}
						\label{TheoremGatheringFarIllustration}
					\end{subfigure}
					\caption[ ]
					{\small Left extremal agent jump (\subref{TheoremGatheringNearIllustration})into/(\subref{TheoremGatheringFarIllustration})over the internal agent interval.} 
					\label{fig:TheoremGathering}
				\end{figure*}
				Hence in all possible cases the span of the gathered agents at the next step never exceeds one.
			\end{proof}
						
			The next theorem demonstrates that the size of internal agents' interval, if bigger than one, will be reduced in finite expected time by one-half of the difference between the interval size and 1.
			We shall then exploit the fact that the number of agents is finite and that the shrinkage can not be infinitesimal, to show that the interval indeed will attain a size less than 1, in finite expected time.
			
			\newpage
			\begin{theorem}\label{thm:FiniteExpectedTimeToHalfShrink}
				Let agents $p_1, p_2, \ldots, p_N$ be initially located at $x_1(0), x_2(0), \ldots, \linebreak x_N(0)$, their behavior being governed by the motion model we consider. Suppose $x_{N-1}(0) - x_2(0) = 1 + S_0$
				for some $S_0 > 0$, i.e. internal agents are not initially gathered inside a unit interval. Let $T = \inf\{t~:~x_{N-1}(t) - x_2(t) \leq 1 + \frac{S_0}{2}\}$ - be the first time, when all the
				internal agents are inside an interval bounded by $1 + \frac{S_0}{2}$, then 
				\begin{equation*}
					\expectation{T} < \frac{1}{1-2\varepsilon} \left( (N-2) \left\lceil \frac{S_0}{2} \right\rceil + \left(x_N(0) - x_1(0) - 1\right)\right).
				\end{equation*}
			\end{theorem}
			\begin{proof}
				Locate two fictional \enquote{beacon agents} $p_F^L$ and $p_F^R$ at the locations defined as follows:
				\begin{enumerate}[label=(\alph*)]
					\item
						$p_F^L$ at $x_F^L(0) = x_2(0) + \frac{S_0}{2}$
					\item
						$p_F^R$ at $x_F^R(0) = x_{N-1}(0) - \frac{S_0}{2}$
				\end{enumerate}
				Obviously, $x_F^R(0) - x_F^L(0) = 1 + S_0 - 2\frac{S_0}{2} = 1$. \\
				
				Now consider the agents to the right of $x_F^R(0)$ and the action of $p_R$ and the agents to the left of $x_F^L(0)$ and the action in time by $p_L$. Clearly there will be \textbf{no interaction}
				between the two dynamic processes to the left and to the right of the interval $\left[x_F^L(0), x_F^R(0)\right]$ until one of the agents $p_R$ or $p_L$ will fully sweep
				all agents located in either the interval $(-\infty, x_F^L(0))$ or in the interval $(x_F^R(0), \infty)$, into the unit interval $\left[x_F^L(0), x_F^R(0)\right]$.
				Indeed no agents from the left can cross into the right region until all of them \enquote{jumped the fence} at $x_F^L(0)$ and the same happens in the opposite direction! \\
				
				\noindent
				Therefore we have that in a finite expected time upper bounded by
				\[
					\frac{1}{1-2\varepsilon} \left( (N-2) \left\lceil \frac{S_0}{2} \right\rceil + \left(x_N(0) - x_1(0) - 1\right)\right)
				\]
				the span of the \enquote{internal}, non-mobile agents will shrink to be at most $1 + \frac{S_0}{2}$. \\
				
				The bound is explained as follows : if we denote by $T_L$ - a random time it takes the agents left of $x_F^L(0)$ to \enquote{jump the fence} and by $T_R$ - the random time it takes
				the agents right of $x_F^R(0)$ to \enquote{jump the fence}, then clearly $T$, the first moment when one of the $\frac{S_0}{2}$ intervals will be cleared of agents is bounded above by $\min \{T_L, T_R \}$. 
				We have then, that in the worst case, we will need at most all internal agents to be swept a distance of at most $\left\lceil \frac{S_0}{2} \right\rceil$, and also an extremal one must move all the 
				way to reach the fence. Hence $\expectation{T = \min\{T_L, T_R\}} < \expectation{\text{worst extremal excursion time}}$, which is the expression above.
			\end{proof}
			
			We next prove the following simple fact
			\begin{lemma}\label{S1PointsLemma}
				Let $x_1, x_2, \ldots x_n$ be a set of real numbers, such that $\{x_i\} \neq \{x_j\}$ for all $i \neq j$ (i.e. their fractional parts are all different). Define 
				$$d \vcentcolon= \min\limits_{i \neq j} \{\abs{\{x_i\} - \{x_j\}}, 1 - \abs{\{x_i\} - \{x_j\}}\}$$
				Then, if for some $i$, $j$ $~\abs{x_i - x_j} > 1$, we must have that: $\abs{x_i - x_j} \geq 1 + d$.
			\end{lemma}
			\begin{proof}
				Write $x_i = s_i + r_i$, where $r_i \in [0, 1)$ and $s_i \in \mathbb{Z}$. Then $\abs{x_i - x_j} = \abs{(s_i-s_j) + (r_i-r_j)}$ and $-1 < r_i-r_j < 1$.
				
				If $x_i - x_j > 1$, then two cases are possible: 
				\begin{itemize}
					\item
						$r_i > r_j$, then $x_i - x_j = (s_i-s_j) + (r_i-r_j)$, $s_i - s_j \geq 1$ and $r_i-r_j = \abs{r_i-r_j} = \abs{\{x_i\}-\{x_j\}} \geq d$. This yields $x_i - x_j \geq 1+d$.
					\item
						$r_i < r_j$, then $x_i - x_j = (s_i-s_j - 1) + (1 - (r_j-r_i))$, $(s_i - s_j - 1) \geq 1$ and $(1 - (r_j-r_i)) = 1 - \abs{r_j-r_i} = 1 - \abs{\{x_j\}-\{x_i\}} \geq d$. 
						This again yields $x_i - x_j \geq 1+d$.
				\end{itemize}
				
				In case $x_i - x_j < -1$, it follows $x_j - x_i > 1$ and we apply previous argument by exchanging roles of indexes $i$ and $j$. Hence in both cases the claim follows.
			\end{proof}
			
			\textbf{Note}, if $\{x_i(0)\}$ are the fractional parts of the initial locations of the agents on the line, then these fractional parts are invariant under the evolution process 
				since agents jump unit steps.

			\noindent
			Assuming, as we do, that all initial fractional parts are distinct we have the following result:
			Define $d$ as in \autoref{S1PointsLemma} to be the smallest fractional difference of all the initial agent pair locations. If $x_{N-1}(t) - x_2(t) > 1$, then necessarily
			$x_{N-1}(t) - x_2(t) \geq 1 + d$.
			
			\begin{theorem}\label{thm:DroppingUnder1}
				In the setting of \autoref{thm:FiniteExpectedTimeToHalfShrink}, let $T = \inf\{t~:~x_{N-1}(t) - x_2(t) \leq 1\}$, i.e. the first time when all the
				internal agents are inside an interval bounded by $1$, then 
				\begin{equation*}
					\expectation{T} < \frac{1}{1-2\varepsilon} (N \cdot(S_0 + \ceil{\log_2 \frac{S_0}{d}}) + (x_N(0) - x_1(0) - S_0 - 1) ).
				\end{equation*}
			\end{theorem}
			\begin{proof}
				From the \autoref{thm:FiniteExpectedTimeToHalfShrink}, given that at time $0$, $(x_{N-1}(0) - x_2(0)) = 1 + S_0$, we have at a random time $T_1$ with finite expectation that 
				$(x_{N-1}(T_1) - x_2(T_1)) \leq 1 + \frac{S_0}{2}$.
				We next consider the process with the constellation of agents at the moment where one of the active extremal agents cleared out an interval of length $\frac{S_0}{2}$ 
				on one side of the span of \enquote{internal agents}. At this moment ($T_1$, the initial time for the next phase) all internal agents are spanning an interval of length at most $1 + \frac{S_0}{2}$.
				Therefore by \autoref{thm:FiniteExpectedTimeToHalfShrink}, after a random time span of $T_2$, again having finite expectation, we find the internal points gathered within 
				an interval of $1 + \frac{S_0}{4}$, etc.
				
				After $k$ such steps, each with finite expected duration, we shall find the internal agents within an interval of length at most $1 + \frac{S_0}{2^k}$. The decrease of the upper bound value 
				on the span of internal agents at step $k$ will be at least $\frac{S_0}{2^{k+1}}$. Recall now that $d$ is the smallest fractional difference of all possible agent pair locations. Suppose at step $k_f$
				(at time $T^* \vcentcolon= T_1 + T_2 + \ldots + T_{k_f}$), we attain for the first time
				$\frac{S_0}{2^{k_f}} < d$ but, still we have $x_{N-1}(T^*) - x_{2}(T^*) > 1$. By \autoref{S1PointsLemma} we must have
				\begin{equation*}
					x_{N-1}(T^*) - x_{2}(T^*) \geq 1 + d
				\end{equation*}
				However, since
				\begin{equation*}
					x_{N-1}(T^*) - x_{2}(T^*) \leq 1 + \frac{S_0}{2^{k_f}} < 1 + d
				\end{equation*}
				leads to a contradiction, we must have an interval $x_{N-1}(T^*) - x_{2}(T^*) \leq 1$ and $T \leq T^*$.
				This proves that, at some step before $k_f = \ceil{\log_2 \frac{S_0}{d}}$ all the internal points will be gathered in an interval of unit length.
				
				Using the upper bound for every $T_1, T_2, \ldots$ we obtain
				\begin{equation*}
					\begin{array}{r c l}
						\expectation{T}
						& \leq & \expectation{T_1} + \expectation{T_2} + \ldots + \expectation{T_{k_f}} \\ [9pt]
						& \leq & \frac{N \left\lceil \frac{S_0}{2} \right\rceil}{1-2\varepsilon} + \frac{N \left\lceil \frac{S_0}{4} \right\rceil}{1-2\varepsilon} 
						+ \ldots + \frac{N \left\lceil \frac{S_0}{2^{k_f}} \right\rceil}{1-2\varepsilon} + \Delta \\ [9pt]
						& \leq & \frac{N (S_0 + k_f)}{1-2\varepsilon} + \Delta
					\end{array}
				\end{equation*}
				
				\noindent
				We still need to evaluate $\Delta$ here. Starting at each time $T_1, T_2, \ldots$ extremal agents need to sweep by their biased walk distances of
				$\left\lceil \frac{S_0}{2} \right\rceil, \left\lceil \frac{S_0}{4} \right\rceil, \ldots$ respectively, with the exception of the first interval $T_1$, when an additional initial \enquote{gap}
				had to be traversed by one of  extremal agents. The possible initial \enquote{gaps} were, $x_2(0) - x_1(0)$ and $x_N(0) - x_{N-1}(0)$. For the upper bound we take the initial traversal length to be the 
				sum of these quantities. After reordering we have for $\Delta \nobreak= \frac{(x_N(0) - x_1(0)) - 1 - S_0}{1-2\varepsilon}$, hence we obtain
				\begin{equation*}
					\expectation{T} < \frac{N \cdot(S_0 + \ceil{\log_2 \frac{S_0}{d}}) + (x_N(0) - x_1(0) - S_0 - 1) }{1-2\varepsilon}
				\end{equation*}
				
			\end{proof}
			
			\noindent
			To summarize, we have the following results so far:
			\begin{itemize}
				\item
					Consider the span of the non-extremal agents' constellation at time $t=0$ on $\mathbb{R}$ as
					\begin{equation*}
						L(0) \triangleq x_{N-1}(0) - x_2(0) \triangleq 1 + S_0
					\end{equation*}
					and with $S_0 > 0$.
					Due to the actions of the \enquote{erratic extremist} agents, while the span of the \enquote{core} agents is greater than $1$ (i.e. it is $L(t) = 1 + S$, with $S > 0$), 
					we have that $x_2(t)$, the location of the second agent in the reordered naming of agents, can only increase, and similarly $x_{N-1}(t)$ can only decrease.					
					Hence, while $L(t)$ is bigger than one, it will be a non-increasing sequence in time. In finite expected time $L(t)$ becomes less than $1$ 
					and the subsequent actions of the extremists can never make it exceed $1$.	
				\item \label{SummaryResult3}
					Following the gathering of the \enquote{core} agents to a consensus interval less than 1 after a finite expected time, the total distance between $p_1$ and $p_N$ will be a sum of three parts : 
					the interval occupied by \enquote{core} agents of size at most $1$ and two distances from the consensus \enquote{core} interval to the left and right \enquote{extremists}.
			\end{itemize}
			
		\newpage
		\subsection{The total span of agents after gathering}\label{Result4}
			In \autoref{Result3} we provided a bound of $\nicefrac{\varepsilon}{(1-2\varepsilon)}$ on the expected length of maximal excursions of an extremal agent from a fixed point. 
			Since expectation is linear we can provide a rough bound on the total span of agent locations as the sum of $1$, (which upper bounds the span of the gathered \enquote{core}, or consensus agents)
			and expected maximal excursions to the left and right made by the \enquote{extremist} agents. This argument yields, roughly
			\begin{equation*}
				\expectation{x_N(t) - x_1(t)} \leq 1 + \frac{2\varepsilon}{1-2\varepsilon}
			\end{equation*}
			The Markov's inequality ($\forall a > 0~~\probability{X \geq a} \leq \frac{\expectation{X}}{a}$) then provides
			\begin{equation}\label{eq:EasyBound}
				\probability{x_N(t) - x_1(t) \geq k} \leq \frac{1}{k} + \frac{2\varepsilon}{k}\cdot\frac{1}{1-2\varepsilon} \approx \frac{1}{k}
			\end{equation}
			Therefore, we have qualitatively that $\probability{x_N(t) - x_1(t) \in [k, k+1]} = \Theta(\frac{1}{k^2})$.
			
			\noindent
			However, we can do even better.
			
			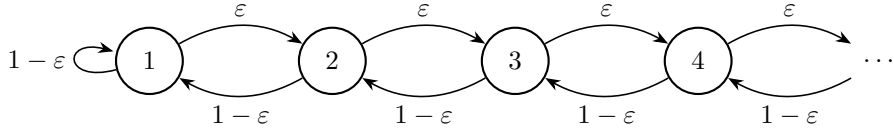
\begin{figure}[h!b]
				\input{images/RPMarkovChain}
				\caption{Left-biased bounded random walk used to bound extremal agent distance from the internal core agents}
				\label{fig:MarkovChain}
			\end{figure}
			
			\noindent
			Let us introduce a left-biased, partially reflective and bounded-from-the-left random walk on the state space $\{1, 2, \ldots \}$ (see \autoref{fig:MarkovChain}). Further, consider each state as
			representing the extremal agent's current distance from the farthest internal agent rounded to the closest bigger integer. The probability to move right, i.e. away from the \enquote{core} (which is
			the gathered, internal agents span), at every state is $\varepsilon$, and the probability to move closer to the \enquote{core} is $(1-\varepsilon)$. The right extremal agent can, with probability $(1-\varepsilon)$,
			jump to the left, but at state $1$ such a jump constitutes a move over all the internal agents. In this case, a new extremal agent \enquote{emerges}, maintaining the distance from the farthest left internal agent 
			just below $1$ (e.g. \autoref{TheoremGatheringToUnitInterval}). This can happen in two ways. Either the other extremal agent jumps over the \enquote{core}, or the closest internal agent becomes \enquote{exposed} 
			and turns into the right extremal one.
			
			After the convergence of the internal agents, suppose we couple the (right) extremal agent's moves to the above defined random walk, i.e. the random walk proceeds exactly following the decisions of extremal agent.
			
			\begin{claim}
				The random walk defined above provides an upper bound on the distance of the extremal agent (at $x_N(t)$) from the farthest internal agents (at $x_2(t)$).
			\end{claim}
			\begin{proof}
				Let $X(t)$ denote the state of random walk at time $t$. Suppose at time $t = T$, $X(T) \geq x_N(T) - x_{2}(T)$, i.e. random walk is at state `at least the distance of right extremal agent from 
				the farthest internal agent'.
				At $t=T+1$ one of the following things can happen. 
				\begin{itemize}
					\item
						The right extremal agent decides to jump right. In such a case, the distance to the extremal agent increases by at most $1$, which corresponds to an increase in the random walk
						position. 
						\begin{equation*}
							\begin{array}{l l l l l}
								X(T+1) & = & X(T) + 1 & \geq & (x_N(T) + 1) - x_2(T) \\
								& = & x_N(T + 1) - x_2(T) & \geq & x_N(T + 1) - x_2(T + 1)
							\end{array}
						\end{equation*}
						The last inequality follows from the fact, that the left-most internal agent can only \enquote{move} to the right, due to the action of the left \enquote{extremist}.
					\item
						The right extremal agent decides to jump left, but remains the right extremal agent at $t = T+1$, and $x_N(T) - x_{2}(T) > 1$. Therefore,
						\begin{equation*}
							\begin{array}{r c l l l}
								X(T+1) & = & X(T) - 1 & \geq & (x_N(T) - 1) - x_2(T) \\
								& = & x_N(T + 1) - x_2(T) & \geq & x_N(T + 1) - x_2(T + 1)
							\end{array}
						\end{equation*}
						The last inequality is explained as in the preceding case.
					\item
						The right extremal agent decides to jump left, stops being the right extremal agent at $t = T+1$, and $x_N(T) - x_{2}(T) > 1$.
						We assumed, that $X(T) \geq x_N(T) - x_2(T)$ which is equivalent to $X(T) \geq 2$, hence by definition of coupling $X(T+1) \geq 1$. Two situations are possible: 
						the internal agent at $x_{N-1}(T)$ \enquote{emerged} to be the 
						right extremal agent at time $T+1$ or the left extremal agent at time $T+1$ jumps over all the other agents to the right and becomes the right extremal one.
						We have $x_2(T+1) \geq x_2(T)$, since the right extremal agent becomes the internal agent. Also $x_N(T+1) = x_1(T) + 1 \leq x_2(T) + 1$, since only the extremal agents actually move.
						In both cases it follows that
						\begin{equation*}
							x_N(T+1) - x_2(T+1) \leq (x_2(T) + 1) - x_2(T) = 1 \leq X(T+1)
						\end{equation*}
					\item
						The right extremal agent decides to jump left and $x_N(T) - x_{2}(T) \leq 1$. In such a case $X(T) \geq 1$ and $X(T+1) \geq 1$, because $1$ is the lowest value the random walk could attain. 
						The distinctive difference from the previous case is that the right extremal agent moves over all the internal agents. We have then three cases. The first is when the right extremal agent becomes 
						the left-most internal agent, hence 
						\begin{equation*}
							x_2(T+1) = x_N(T) - 1 \geq x_{N-1}(T) - 1 = x_N(T+1) - 1.
						\end{equation*}
						In the second and third cases, it becomes the left extremal agent. We differentiate between those two cases considering the new role of the previous left extremal agent.
						If it becomes a new right extremal agent, we have
						\begin{equation*}
							x_N(T+1) = x_1(T) + 1 \leq x_2(T) + 1 = x_2(T + 1) + 1.							
						\end{equation*}
						Otherwise,
						\begin{equation*}
							\begin{array}{r c l l l}
								x_N(T+1) & = & x_{N-1}(T) & \leq & (x_N(T) - 1) + 1 \\
								& = & x_1(T + 1) + 1 & \leq & x_2(T + 1) + 1
							\end{array}
						\end{equation*}
						
						In all the above cases, we conclude $x_N(T+1) - x_2(T+1) \leq 1 \leq X(T + 1)$, as claimed.
				\end{itemize}
			\end{proof}
			
			Returning to analyze the \enquote{upper bounding} random walk we have the following:	if $\varepsilon < \nicefrac{1}{2}$ the above random walk is positive recurrent, and aperiodic, 
			hence has a stationary distribution $\pi$ that is determined by the balance equations
			\begin{equation*}
				\varepsilon \cdot \pi(k) = (1 - \varepsilon) \cdot \pi(k+1)
			\end{equation*}
			Along with the normalization condition $\sum\limits_{k=1}^{\infty} \pi(k) = 1$, this provides the steady state distribution $\pi = [\pi(1)~~\pi(2)~~\ldots]$ with
			\begin{equation} \label{eq:LongTermDistributionOfExtremalAgentsDistances}
				\pi(k) = \left(\frac{\varepsilon}{1-\varepsilon}\right)^{k-1}\frac{1-2\varepsilon}{1-\varepsilon}~~~\forall k \in \{1, 2, \ldots \}.
			\end{equation}
			
			The above analysis is symmetrically applicable to the random walk of the left extremal agent.			
			We then have two independent and identically distributed walks, upper bounding the distance of the right and left extremal agents from the \enquote{core's} left and right boundaries. 
			Denoting them by $X(t)$ and $Y(t)$, we have
			\begin{equation*}
				x_N(t) - x_1(t) \leq (x_N(t) - x_2(t)) + (x_{N-1}(t) - x_1(t)) = X(t) + Y(t)
			\end{equation*}
			
			Here we are interested in assessing $\probability{x_N(t) - x_1(t) \leq k}$, hence we can estimate a lower bound for $\probability{x_N(t) - x_1(t) \leq k}$ by $\probability{X(t) + Y(t) \leq k}$. Therefore, consider
			\begin{equation}\label{eq:DiscreteConvolutionEquation}
				\probability{X+Y \geq k} = \sum\limits_{i=1}^{k-2}\probability{X=i}\probability{Y \geq k-i} + \probability{X \geq k - 1}
			\end{equation}
			
			In the steady state we have that $\probability{X=k}$ is just a $\pi(k)$, and $\probability{X \geq k} = \sum\limits_{i=k}^{\infty}\pi(i)$. Therefore,
			\begin{equation*}
				\sum\limits_{i=k}^{\infty}\pi(i) = \sum\limits_{i=k}^{\infty}\left(\frac{\varepsilon}{1-\varepsilon}\right)^{i-1}\frac{1-2\varepsilon}{1-\varepsilon}
				= \left(\frac{\varepsilon}{1-\varepsilon}\right)^{k-1}
			\end{equation*}
			Thus, we obtained the following simple expression
			\begin{equation*}
				\probability{X \geq k} = \left(\frac{\varepsilon}{1-\varepsilon}\right)^{k-1}
			\end{equation*}
			Using this result in \eqref{eq:DiscreteConvolutionEquation} produces for $k$ greater than two
			\begin{equation*}			
				\probability{X+Y \geq k} = \sum\limits_{i=1}^{k-2}\left(\frac{\varepsilon}{1-\varepsilon}\right)^{i-1}\frac{1-2\varepsilon}{1-\varepsilon}\cdot\left(\frac{\varepsilon}{1-\varepsilon}\right)^{k - i - 1}
				+ \left(\frac{\varepsilon}{1-\varepsilon}\right)^{k-2},
			\end{equation*}			
			which after few algebraic manipulations provides
			\begin{equation*}			
				\probability{X+Y \geq k} = \left(\frac{\varepsilon}{1-\varepsilon}\right)^{k-2}\left((k-2)\cdot\frac{1-2\varepsilon}{1-\varepsilon}+1\right).
			\end{equation*}
			
			\noindent
			We summarize these findings as follows :
			\begin{theorem}\label{thm:IntervalDistributions}
				After the internal agents gather in an interval of length 1, the distribution of interval lengths' containing all the agents is lower bounded by
				\begin{equation}\label{eq:LongTermDistributionOfExtremalAgentDistances}
					\probability{x_N(t) - x_1(t) < k} \leq \probability{X+Y < k} \approx 1 - k\left(\frac{\varepsilon}{1-\varepsilon}\right)^{k-2}.
				\end{equation}
			\end{theorem}
	
		\subsection{On arbitrary initial position of agents}
			In proving \autoref{thm:DroppingUnder1} we have assumed all the agents locations' fractional parts are different. We can slightly change the model
			to accommodate for cases in which some agents may share the same location. Of course, the problem arises when several agents find themselves sharing extremal locations. In such cases their motions
			must be specified and disambiguated. Suppose several agents share the same place and all other agents are located on exactly one side either to the left, or to the right. 
			We assume that only one of these extremal agents will become \enquote{erratic}, and move at a given time. We can then readily prove a claim equivalent to \autoref{thm:DroppingUnder1} in this new model.
			\begin{theorem}
				Let agents $p_1, p_2, \ldots, p_N$ be initially located at $x_1(0), x_2(0), \ldots$, $x_N(0)$, and define 
				\begin{equation*}
					T \vcentcolon= \inf\{t~:~x_{N-1}(t) - x_2(t) \leq 1\}
				\end{equation*} 
				be the first time, when all the internal agents are inside the interval bounded by $1$, then with the modified rule of behavior we have $\expectation{T} < \infty$.
			\end{theorem}
			\begin{proof}
				Since we are not assuming that fractional parts are all different, it is possible that there will be more than one agent  with the same fractional part of their initial (and subsequent)
				locations. Let $\Delta$ be the minimal fractional non-zero distance between two agents.
				\begin{equation*}
					\Delta \vcentcolon= \min\limits_{ \{x_j(0)\} \neq \{x_k(0)\}} \{\{x_j(0) - x_k(0)\}, 1 - \{x_j(0) - x_k(0)\} \},
				\end{equation*}
				In case all the agents share the same fractional part, simply set $\Delta \vcentcolon = 1$.
				
				Step 1. Define a new process with the following initial coordinates: $\linebreak{y_k(0) = x_k(0) + \frac{(k-1)\Delta}{N}}$ for all $k \in \{1, 2, \ldots, N\}$. It is not difficult to see 
				that the \enquote{newly defined locations} $y_1(0), y_2(0), \ldots, y_N(0)$ fulfill the requirements of \autoref{thm:DroppingUnder1}.
				
				Step 2. The \autoref{thm:DroppingUnder1} proves that all agents $(y_k)_{k=1}^n$ gather in expected finite time to the interval of unit length. Denote this time by $T_y$.
				\begin{itemize}
					\item
						By separately handling cases of same and different initial fractional part of the location one can show that for all $k \geq 2$,
						\begin{equation*}
							x_2(t) \leq x_k(t)
						\end{equation*}
						
						In the same manner, for all $k \leq N-1$
						\begin{equation*}
							x_k(t) \leq x_{N-1}(t)
						\end{equation*}
						
						We can then conclude, that all the correspondingly indexed $x$ and their \enquote{shadow $y$-agents} will be called inner and extremal in both models at the same time.
						\newpage
					\item
						Suppose $y_{2}(T_y)$ and $y_{N-1}(T_y)$ are agents which originally had the same fractional part of their respective location. Due to the way, we mapped the coordinates, we know that
						$y_{2}(T_y) < x_2(T_y) + \Delta \leq x_2(T_y) + 1$ and that $y_{N-1}(T_y) \geq x_{N-1}(T_y)$, hence $\abs{x_{N-1}(T_y) - x_2(T_y)} < 2$. 
						But, since $x_2(T_y)$ and $x_{N-1}(T_y)$ have the same fractional part, we conclude that we have $x_{N-1}(T_y) - x_2(T_y) \leq 1$.
					\item
						If $y_{2}(T_y)$ and $y_{N-1}(T_y)$ are not agents which originally had the same fractional part of their respective location, then one of two cases is possible.
						If $y_{N-1}(T_y) < x_2(T_y) + 1$ we have all agents in original model inside the interval $[x_2(T_y), x_2(T_y) + 1)$. Otherwise 
						$x_2(T_y) + 1 \leq y_{N-1}(T_y) \leq y_{2}(T_y) + 1$. But, due to definition of $\Delta$, only points, which have the same fractional part, as $x_2(T_y)$ could fall between $x_2(T_y)+1$ and
						$y_2(T_y)+1$. Hence the latter case is impossible.
				\end{itemize}
				
				\noindent
				It follows, that in all cases $x_{N-1}(T_y) - x_2(T_y) \leq 1$, which implies $T \leq T_y$, and  by \autoref{thm:DroppingUnder1}, $T_y$ has a finite expectation.
			\end{proof}
			
	\vspace*{3\baselineskip}
	\section{Simulation results} \label{Section_Simulations}
		\vspace*{2\baselineskip}
		We next present some simulation results to showcase the validity of the above-presented theoretical predictions.
		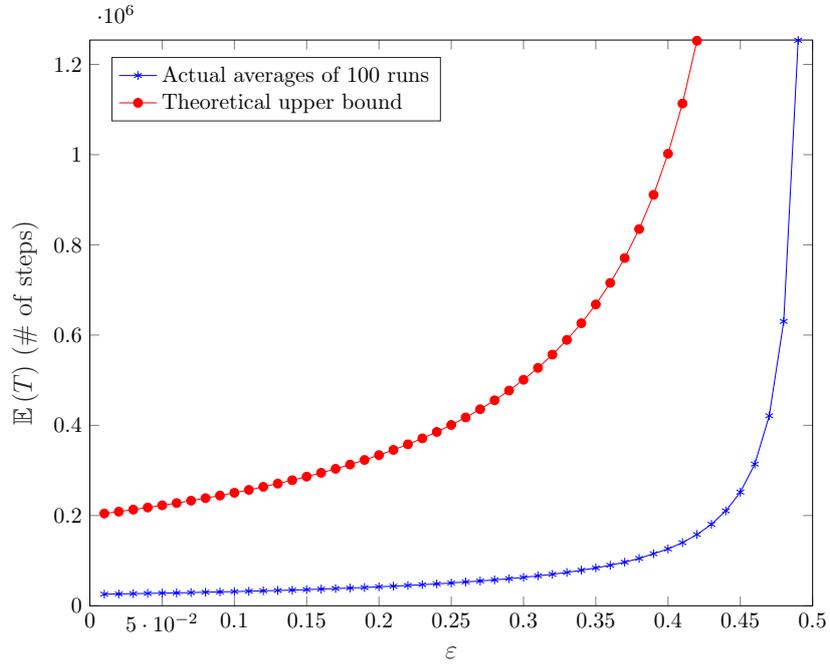
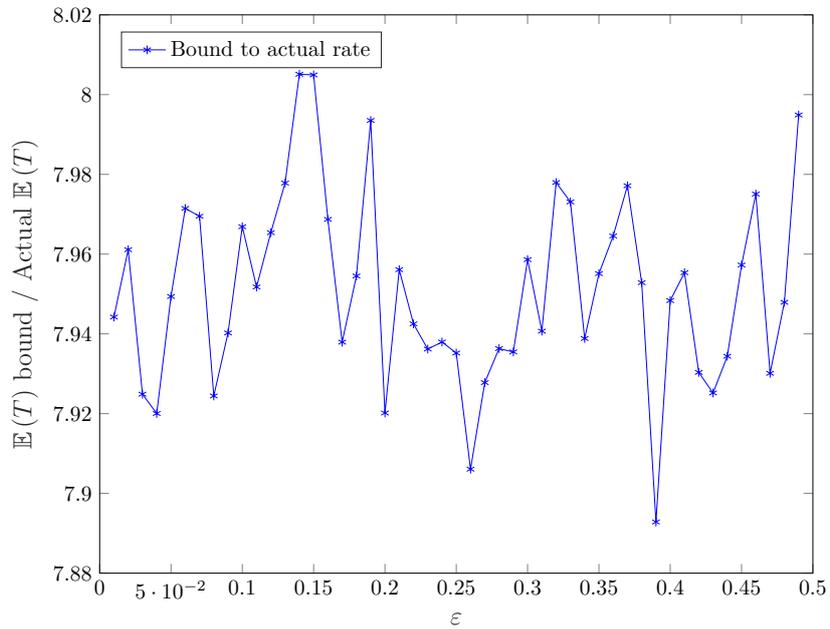
\begin{figure*}        
			\centering
			\begin{subfigure}[b]{0.9\textwidth}  
				\centering 
				\resizebox{\linewidth}{!}{\input{./images/ConvergenceTimeN400S0500.Theoretical.tex}}
				\caption[]%
				{{\small $N = 400$, $S_0 = 500$}}    
				\label{fig:ConvergenceN400S0500}
			\end{subfigure}
			\vskip\baselineskip
			\begin{subfigure}[b]{0.9\textwidth}   
				\centering 
				\resizebox{\linewidth}{!}{\input{./images/PredictionRateN400S0500.tex}}
				\caption[]%
				{{\small $N = 400$, $S_0 = 500$}}    
				\label{fig:PredictionRateN400S0500}
			\end{subfigure}
			\caption[ ]
			{\small (\subref{fig:ConvergenceN400S0500}) Convergence times as a function of probability of motion in the wrong direction($\varepsilon$), $N = 400$, $S_0 = 500$. (\subref{fig:PredictionRateN400S0500}) 
			Theoretical upper bound to measured convergence time ratio vs $\varepsilon$. Each point on the actual results' line is an average of 100 different simulations with the same parameters set.} 
			\label{fig:ConvergenceEpsilon2}
		\end{figure*}
		In 
		\autoref{fig:ConvergenceEpsilon2} we present the result of simulation runs with a different values for $\varepsilon$ and fixed number of agents $N$ located at random points uniformly distributed in an 
		initial interval of size $1 + S_0$. The simulations measured the time to convergence of the inner agents to an interval of length one. The theory predicts that the expected time to gathering is bounded as follows
		\begin{equation*}
			\expectation{T} \leq \frac{N \cdot(S_0 + \ceil{\log_2 \frac{S_0}{d}}) + (x_N(0) - x_1(0) - S_0 - 1) }{1-2\varepsilon}
		\end{equation*}
		As predicted, the average convergence times exhibit a hyperbolic dependence on $\varepsilon$.
		\begin{figure*}        
			\centering
			\begin{subfigure}[b]{0.9\textwidth}
				\centering
				\resizebox{\linewidth}{!}{\input{./images/ConvergenceTimeEpsilon0.1N400.Theoretical.tex}}
				\caption[Convergence times. $N = 400$, $\varepsilon = 0.1$]%
				{{\small $N = 400$, $\varepsilon = 0.1$}}
				\label{fig:ConvergenceEpsilon01N400}
			\end{subfigure}
			\vskip\baselineskip
			\begin{subfigure}[b]{0.9\textwidth}   
				\centering 
				\resizebox{\linewidth}{!}{\input{./images/PredictionRateEpsilon0.1N400.tex}}
				\caption[]%
				{{\small $N = 400$, $\varepsilon = 0.1$}}
				\label{fig:PredictionEpsilon01N400}
			\end{subfigure}
			\caption[ ]
			{\small (\subref{fig:ConvergenceEpsilon01N400}) Convergence times as a function of initial span ($S_0$). (\subref{fig:PredictionEpsilon01N400}) 
			Ratio between the theoretical upper bound to measured convergence time vs $S_0$. Each point on the actual results' line is an average of 100 different simulations with the same parameter set.} 
			\label{fig:ConvergenceS_0}
		\end{figure*}
		\autoref{fig:ConvergenceS_0} clearly showcases the linear functional dependence between the convergence time and the initial span of internal agents, and implicitly to the initial span of all agents.
		Varying the number of agents $N$ supplies another linear dependency as can be seen in \autoref{fig:ConvergenceN}.
		
		\begin{figure*}        
			\centering
			\begin{subfigure}[b]{0.9\textwidth}  
				\centering 
				\resizebox{\linewidth}{!}{\input{./images/ConvergenceTimeEpsilon0.1S01000.Theoretical.tex}}
				\caption[]%
				{{\small $S_0 = 1000$, $\varepsilon = 0.1$}}    
				\label{fig:ConvergenceEpsilon01S01000}
			\end{subfigure}
			\vskip\baselineskip
			\begin{subfigure}[b]{0.9\textwidth}   
				\centering 
				\resizebox{\linewidth}{!}{\input{./images/PredictionRateEpsilon0.1S01000.tex}}
				\caption[]%
				{{\small $S_0 = 1000$, $\varepsilon = 0.1$}}    
				\label{fig:PredictionEpsilon01S01000}
			\end{subfigure}
			\caption[ ]
			{\small (\subref{fig:ConvergenceEpsilon01S01000}) Convergence times as a function of number of agents ($N$). (\subref{fig:PredictionEpsilon01S01000}) 
			Theoretical upper bound to measured convergence time vs $N$. Each point on the actual results' line is an average of 100 different simulations with the same parameters set.} 
			\label{fig:ConvergenceN}
		\end{figure*}

		\newpage
		In all above experiments we notice that our theoretical bounds are roughly eight times higher than the actual measurements. Recall, that we derived our results based on overly cautious assumptions, 
		namely that one extremal agent is doing the constructive work, while the second extremal agent is randomly wandering outside the interval containing the internal agents. 
		In reality, this is not the case: both agents independently and concurrently contribute to convergence. Hence we should focus on the stochastic process, which is in some sense 
		\enquote{the distance between two independent random walks biased towards each other}. 
		
		Let $X, Y$ be two independent biased random walks with a probability $\varepsilon$ to jump right. For any one of mentioned random walks $\expectation{\text{step length}} = 2\varepsilon - 1$. 
		On the other hand, for the process $Z \triangleq X + Y$, $\expectation{\text{step length of } Z} = 2(2\varepsilon - 1)$. Note, that the \enquote{contraction} process $Z$ describes the distance between two extremal agents,
		with each extremal agent sweeping the internal agents in the direction of its counterpart. Furthermore, on the average, the core convergence will happen approximately around the middle of the initial interval. 
		And we should finally recall the assumed uniform initial spread of agents inside the initial interval at the beginning, implying that each extremal agent will need \enquote{to push} 
		only half of the internal agents. Thus until convergence we have two stochastic processes of the kind we analyzed in this paper, and each starts with half the number of agents, and half the initial interval, 
		and the process will proceed at least twice as fast. Hence, we have 3 factors that each improve by 2 the time to convergence (hence the 8!)
		
		Another aim of our simulations was to asses on the bounds on the total span of agents after gathering.
		\begin{figure*}        
			\centering
			\begin{subfigure}[b]{0.9\textwidth}
				\centering
				\resizebox{\linewidth}{!}{\input{./images/N.TotalDistance.semilogy.tex}}
				\caption[Long-term distribution of span]%
				{{\small Long-term distribution of agent span}}
				\label{fig:TotalDistance}
			\end{subfigure}
			\vskip\baselineskip
			\begin{subfigure}[b]{0.9\textwidth}  
				\centering 
				\resizebox{\linewidth}{!}{\input{./images/N.WBounds.TotalDistanceCDF.semilogy.tex}}
				\caption[]%
				{{\small Cumulative distribution of spans}}    
				\label{fig:TotalDistanceWBounds}
			\end{subfigure}
			\caption[ ]
			{\small (\subref{fig:TotalDistance}) Long-term (steady-state) distribution of agents' total span, (\subref{fig:TotalDistanceWBounds}) Long-term cumulative distribution of total span with lower bounds from
			\autoref{Result3} and \autoref{Result4}. The simulations were done for different values of $N$ and $\varepsilon = 0.1$.} 
			\label{fig:TotalDistances}
		\end{figure*}
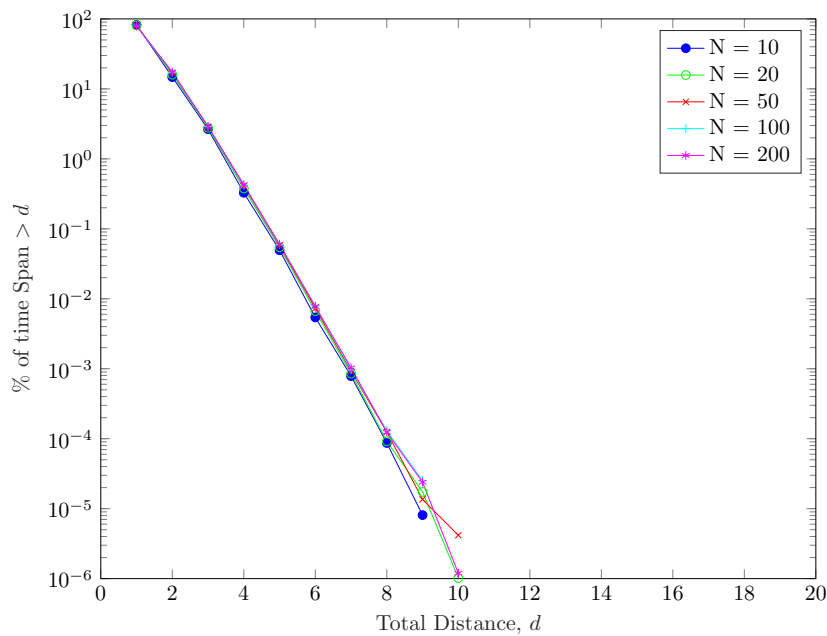
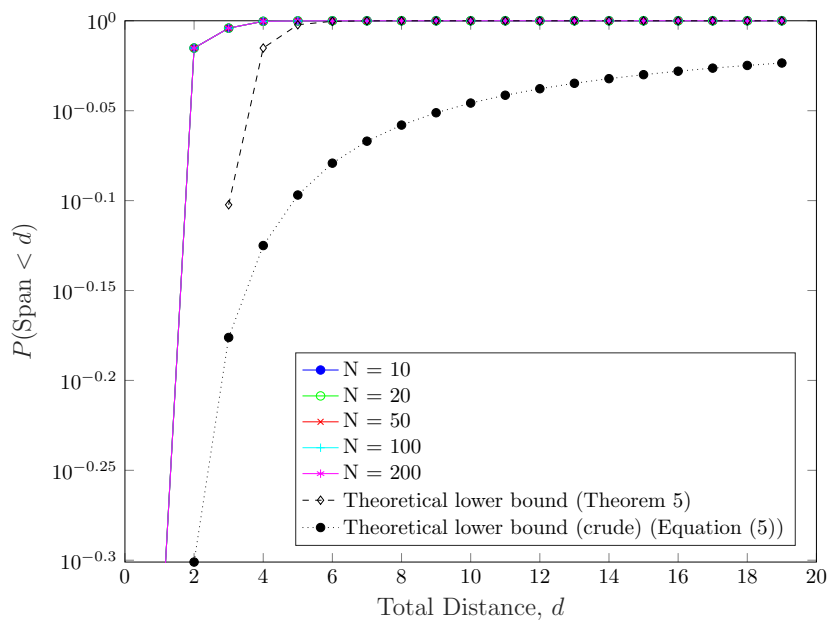
		As can be seen in \autoref{fig:TotalDistances}, using a semi-logarithmic scaling, the probability to find an extremal agent at a specific distance is indeed decreasing exponentially fast,
		according to the bound of \autoref{thm:IntervalDistributions}. The same is true about the results predicted in \autoref{Result4}. In \autoref{fig:TotalDistanceWBounds} we show that application of 
		\autoref{thm:IntervalDistributions} gives a much better bound than the crude evaluation of \eqref{eq:EasyBound}.
		
		\begin{figure*}        
			\centering
			\begin{subfigure}[b]{0.9\textwidth}
				\centering
				\resizebox{\linewidth}{!}{\input{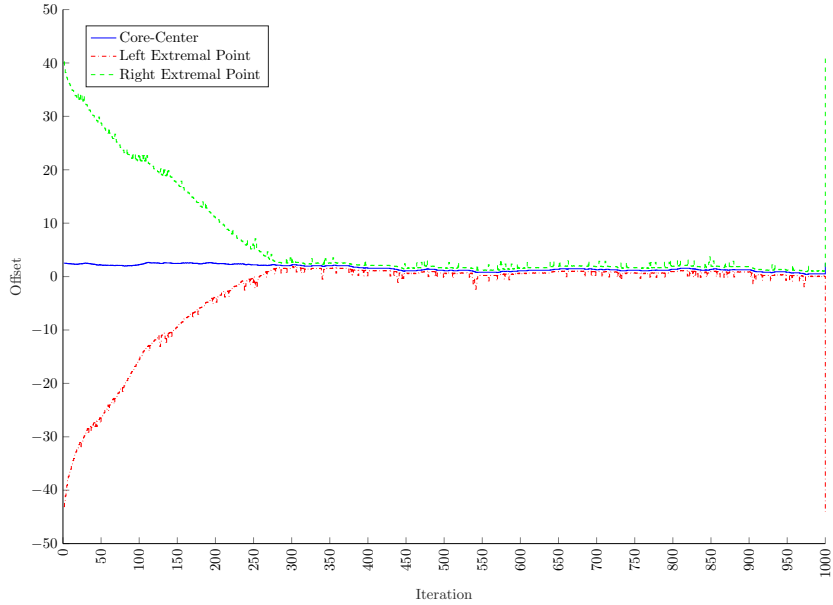}}
				\caption[First 1,000 iterations]%
				{{\small First 1,000 iterations : The gathering process}}
				\label{fig:Centers1}
			\end{subfigure}
			\vskip\baselineskip
			\begin{subfigure}[b]{0.9\textwidth}  
				\centering 
				\resizebox{\linewidth}{!}{\input{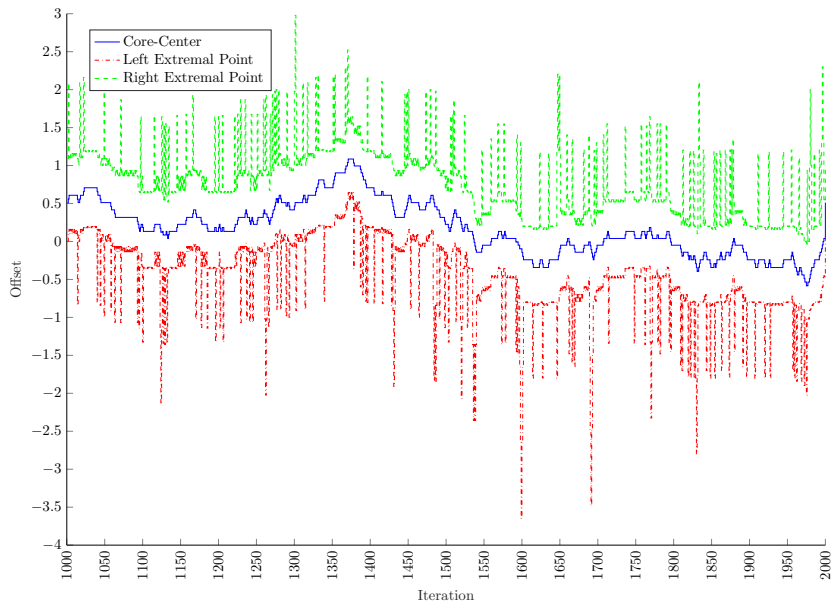}}
				\caption[]%
				{{\small Iterations after gathering ($T$ here was $1000$ and the gathering happened at $t = 300$)}}    
				\label{fig:Centers2}
			\end{subfigure}
			\caption[ ]
			{\small Typical \enquote{core} center and extremal agent location vs. time. From the beginning (\subref{fig:Centers1}) and after the gathering (\subref{fig:Centers2}). Simulations with $N=21$, $\varepsilon=0.1$.	} 
			\label{fig:Centers1000}
		\end{figure*}
		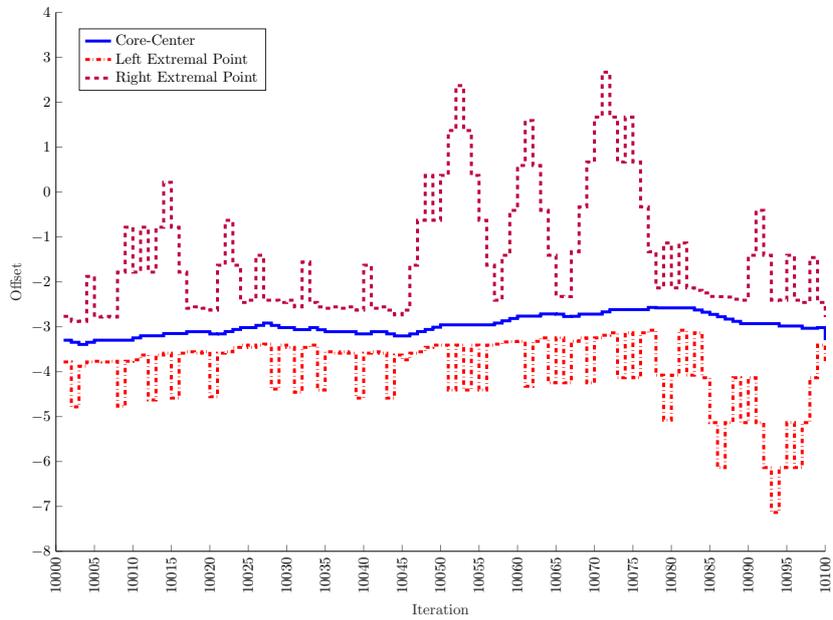
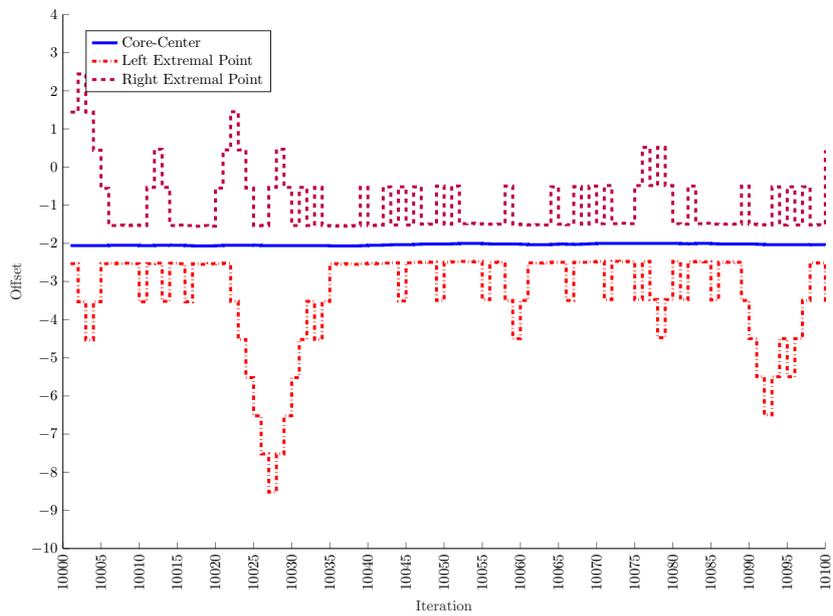
\begin{figure*}        
			\centering
			\begin{subfigure}[b]{0.9\textwidth}   
				\centering 
				\resizebox{\linewidth}{!}{\input{images/centers.3.tex}}
				\caption[]%
				{{\small \enquote{Core} and extremal agents (zoomed). $N = 21$}}
				\label{fig:Centers3}
			\end{subfigure}
			\vskip\baselineskip
			\begin{subfigure}[b]{0.9\textwidth}   
				\centering 
				\resizebox{\linewidth}{!}{\input{images/centers.4.tex}}
				\caption[]%
				{{\small \enquote{Core} and extremal agents (zoomed). $N = 121$}}    
				\label{fig:Centers4}
			\end{subfigure}
			\caption[ ]
			{\small \enquote{Core} center location and both extremal agents vs. time (after gathering, starting from $T=10,000$). Simulations with $\varepsilon=0.3$. Note
			that the \enquote{core} of gathered agents is much more easily moved by \enquote{extremists} when the population is small ($N = 21$)} 
			\label{fig:Centers100}
		\end{figure*}
		
		\begin{figure*}        
			\centering
			\begin{subfigure}[b]{0.9\textwidth}   
				\centering 
				\resizebox{\linewidth}{!}{\input{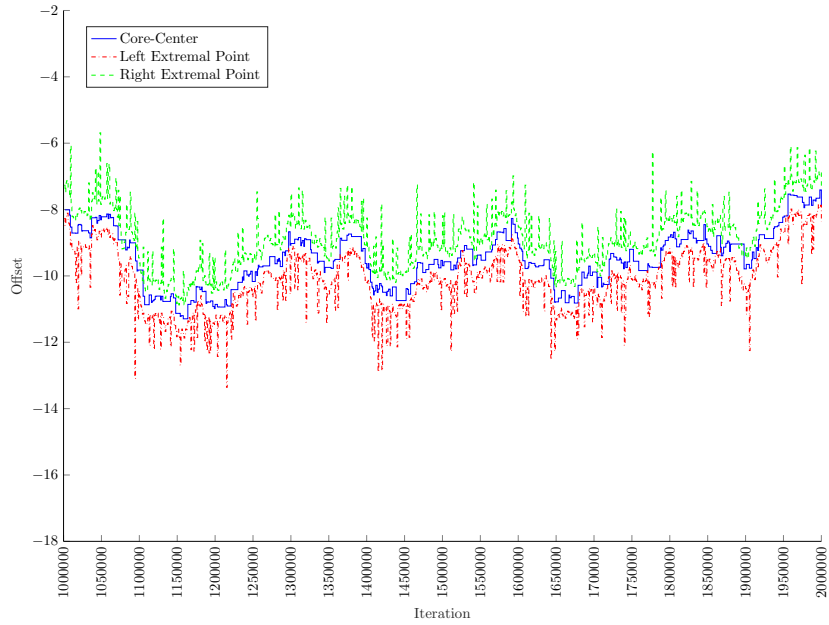}}
				\caption[]%
				{{\small Evolution of \enquote{core} and extremal agents (zoomed out). $N = 200$}}
				\label{fig:Centers5}
			\end{subfigure}
			\vskip\baselineskip
			\begin{subfigure}[b]{0.9\textwidth}   
				\centering 
				\resizebox{\linewidth}{!}{\input{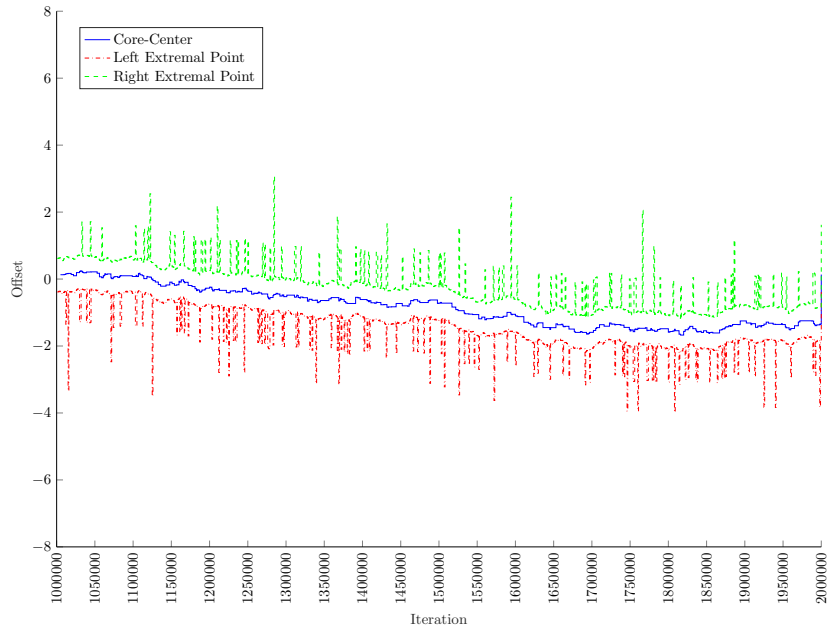}}
				\caption[]%
				{{\small Evolution of \enquote{core} and extremal agents (zoomed out). $N = 1000$}}    
				\label{fig:Centers6}
			\end{subfigure}
			\caption[ ]
			{\small Typical evolution of \enquote{core}'s center of mass location and extremal agents' locations in time (after gathering, starting from $T=1,000,000$. 
				Simulations done for $\varepsilon=0.1$). Note that in both cases the initial center of mass of all agents was at $0$. Note that the \enquote{inertia} of
				society is much higher when $N=1000$ than in case $N=200$.} 
			\label{fig:Centers1000000}
		\end{figure*}
		
		\autoref{fig:Centers1000} presents a typical behavior of the Gathered Core's center of mass and of the two extremal agents. The period before gathering is shown in
		\autoref{fig:Centers1} followed by a display of \enquote{post-gathering} typical behavior in \autoref{fig:Centers2}. Simulations for 
		different number of agents are shown in \autoref{fig:Centers100} and \autoref{fig:Centers1000000}. Unsurprisingly, \Autoref{fig:Centers4, fig:Centers6} prove a much higher inertia of 
		the Core center of mass to the actions of extremal agents, when the number of agents is five times higher.

	\newpage
	\section{Concluding remarks} \label{Section_Conclusions}
		\vspace*{2\baselineskip}
		We here proposed a mathematical model of randomly interacting particles on the line, that could describe opinion dynamics in a society of presumably intelligent agents. 
		Equipped with a simple decision rule, agents eventually get together to a drifting gathered constellation, in finite expected time. All the agents of the system, except two, constitute a core of moderate agents that remain closely
		clustered from that point on. The two \enquote{erratic extremist} agents perform random walks biased toward the \enquote{quasi-stationary} core; once in a while the roles 
		of extremal agents change, when an \enquote{erratic extremist} walker joins the core. We have derived expression for the expected convergence time and the distribution of distances of extremal agents. 
		Computer simulations support our findings.
		
		We believe that the model presented will further help analyze two and higher dimensional models which have a practical importance in a number of areas in multi-agent studies. 
		
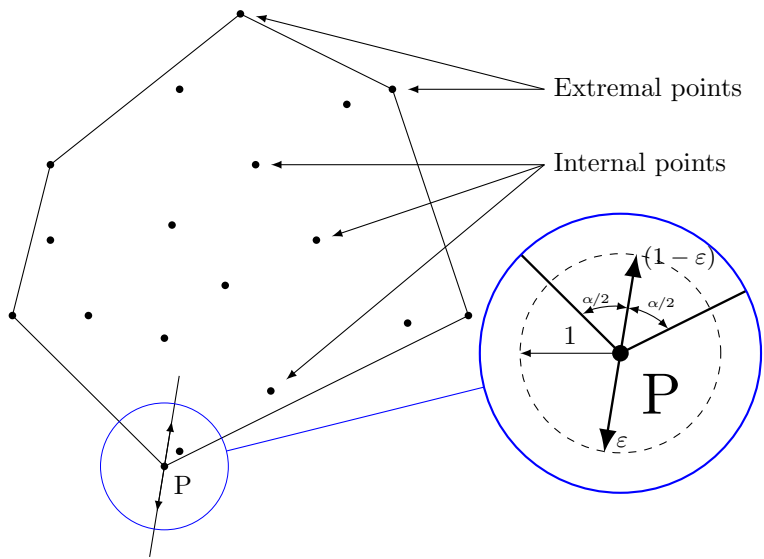
\begin{figure*}[h!]
			\centering
			\input{images/FreddyNewModel}
			\caption{Group of Agents, Convex Hull and zoom on Extremal Agent movement options}
			\label{fig:2DModel}
		\end{figure*}

		An interesting two-dimensional model corresponding to the random evolution process analyzed in this paper could be the following. Assume that the agents' locations are points in the plane $\realnumbers^2$.
		For a group of $N$ agents in the plane the \enquote{extremists} are the ones that define the convex hull of the points. Suppose at each time instant an agent that realizes it is an extreme vertex of the convex hull
		(by sensing the bearing only to all other agents!) decides to move a unit distance along the bisector of the corresponding convex hull angle either toward the other agents (i.e. into the convex hull), with probability
		$(1-\varepsilon)$, or in the opposite direction, with probability $\varepsilon$. (see \autoref{fig:2DModel})
		
\begin{figure*}        
			\centering
			\begin{subfigure}[b]{0.9\textwidth}   
				\centering 
				\resizebox{\linewidth}{!}{\includegraphics{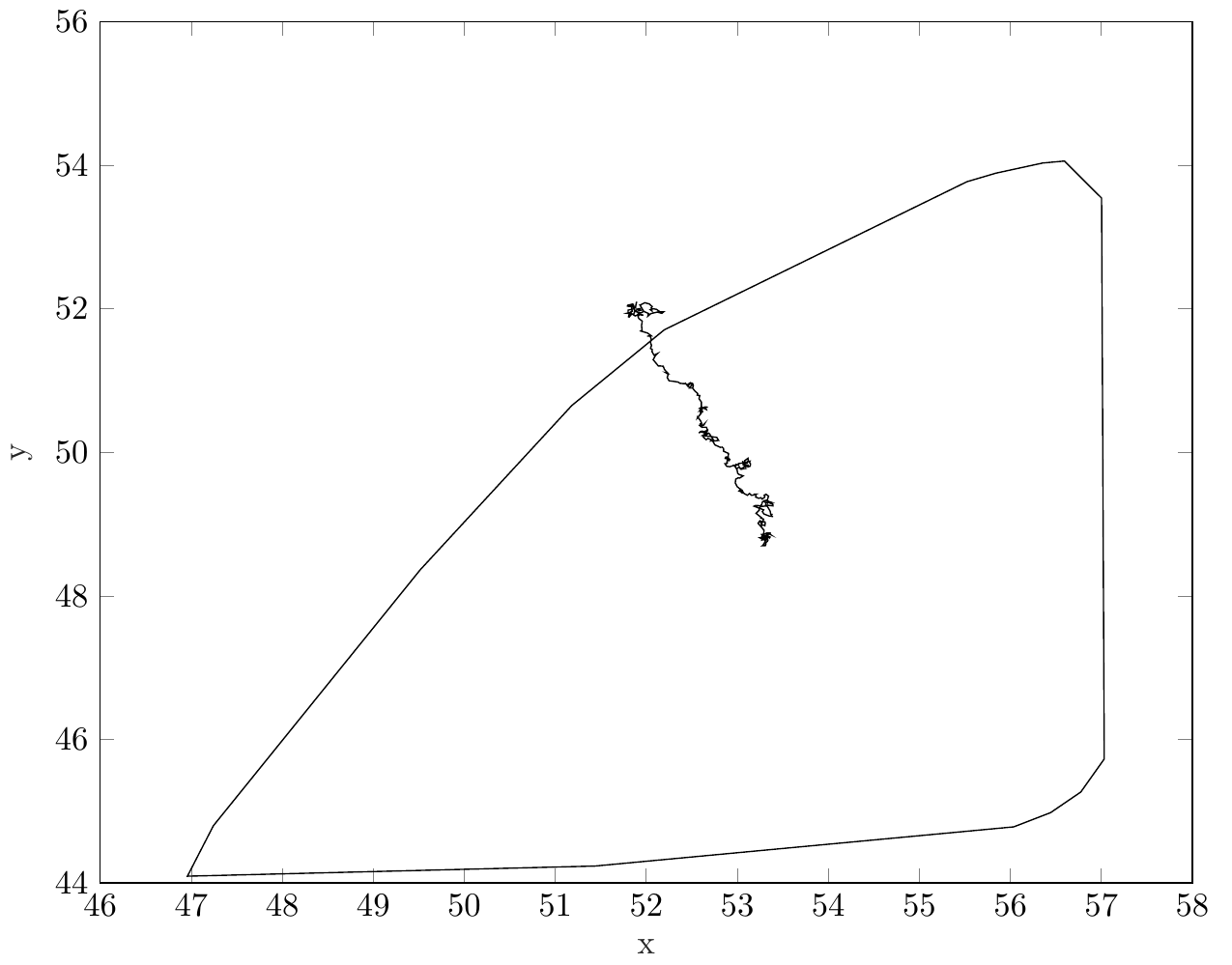}}
				\caption[]%
				{{\small After $400$ iterations}}
				\label{fig:CHCenters400}
			\end{subfigure}
			\vskip\baselineskip
			\begin{subfigure}[b]{0.9\textwidth}   
				\centering 
				\resizebox{\linewidth}{!}{\includegraphics{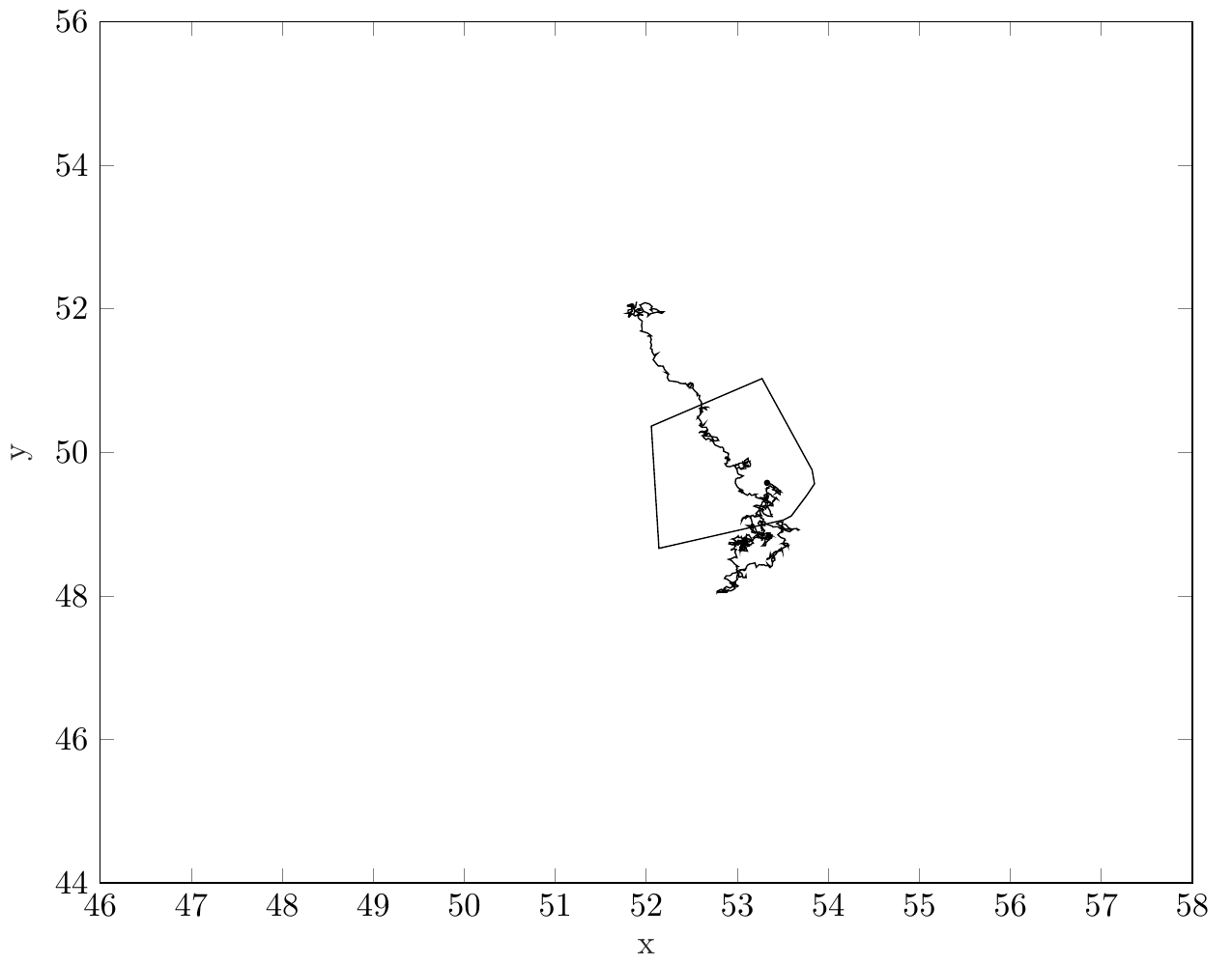}}
				\caption[]%
				{{\small After $1,000$ iterations}}    
				\label{fig:CHCenters1000}
			\end{subfigure}
			\caption[ ]
			{\small Evolution of the Center of Mass of the system and the last Convex Hull after (\subref{fig:CHCenters400}) $400$ and (\subref{fig:CHCenters1000}) $1000$ iterations} 
			\label{2DFreddyCenters}
		\end{figure*}

\begin{figure*}        
			\centering
			\begin{subfigure}[b]{0.9\textwidth}   
				\centering 
				\resizebox{\linewidth}{!}{\includegraphics{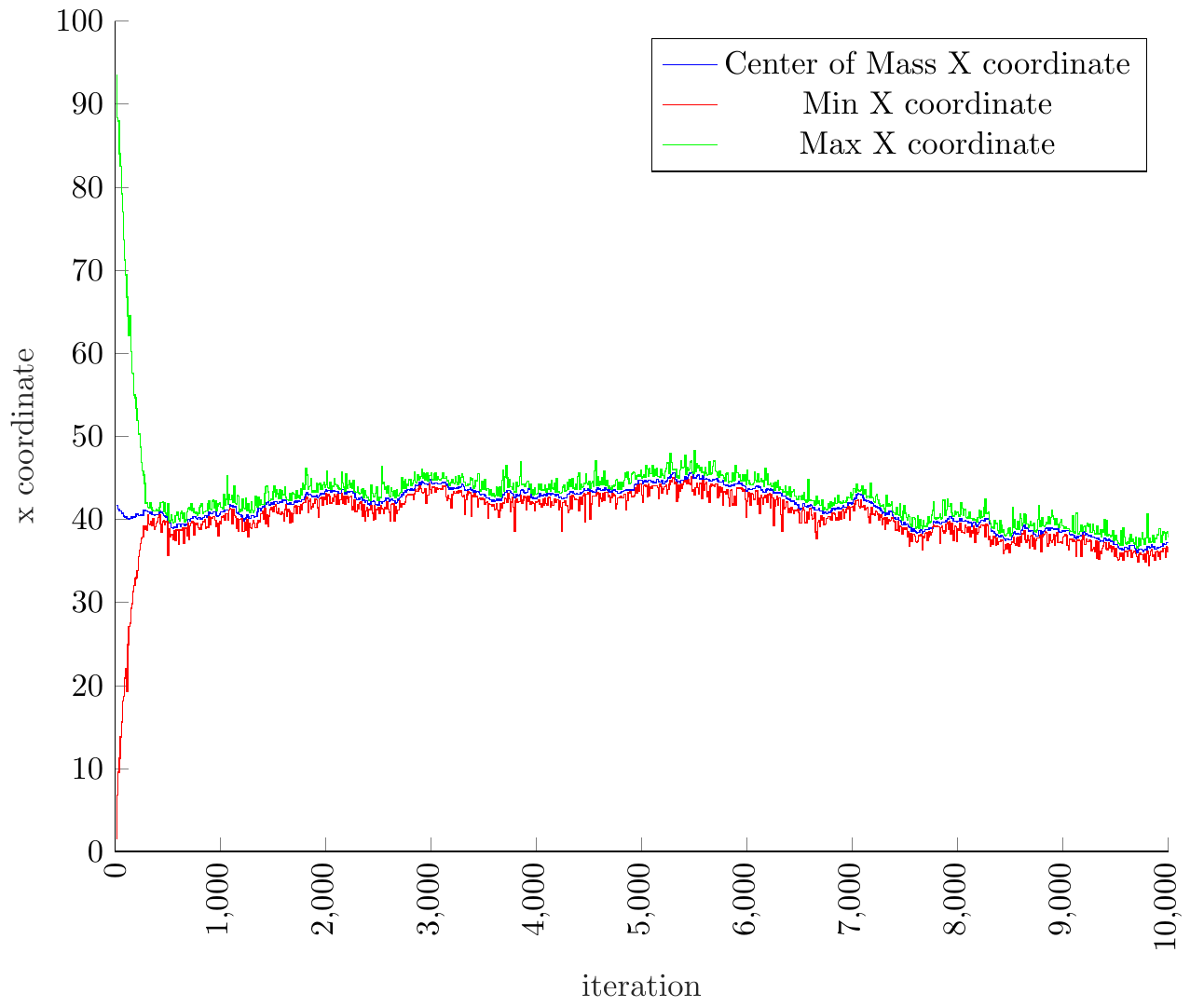}}
				\caption[]%
				{{\small The X direction}}
				\label{fig:ConvexHullX}
			\end{subfigure}
			\vskip\baselineskip
			\begin{subfigure}[b]{0.9\textwidth}   
				\centering 
				\resizebox{\linewidth}{!}{\includegraphics{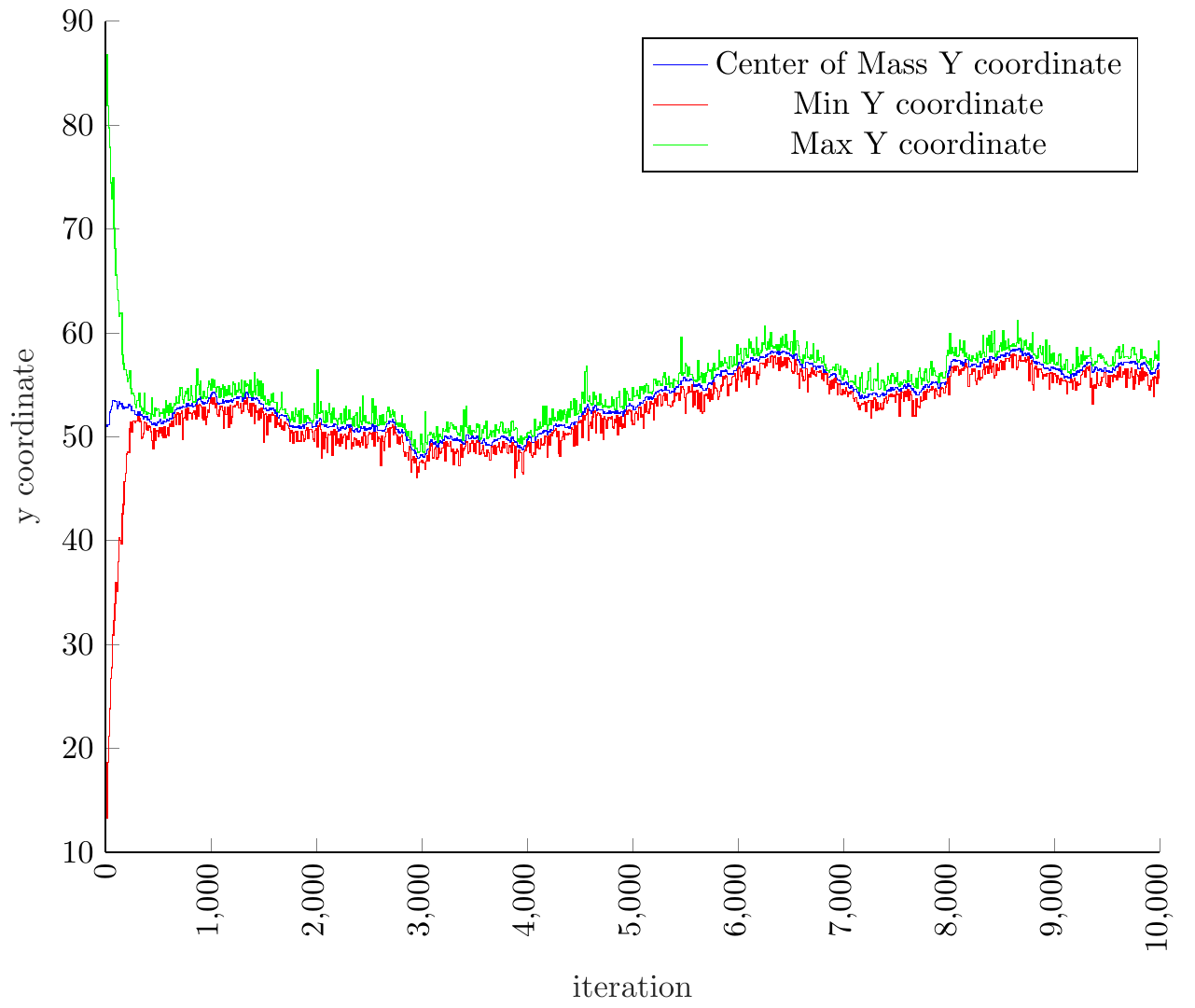}}
				\caption[]%
				{{\small The Y direction}}    
				\label{fig:ConvexHullY}
			\end{subfigure}
			\caption[ ]
			{\small Evolution of the Center of Mass of the system split by (\subref{fig:ConvexHullX}) X direction and by (\subref{fig:ConvexHullY}) Y direction} 
			\label{2DFreddyCentersSplit}
		\end{figure*}

		\newpage
		Preliminary simulations with this model show that indeed the population gathers to a small region in the plane (see \autoref{2DFreddyModelEvolution}) and the gathered group performs a random walk in the plane 
		(\Autoref{2DFreddyCenters, 2DFreddyCentersSplit}). 
		We plan to study this and several variations of such models in the near future.
		
		\vspace*{7\baselineskip}
		\input{images/FreddyConvexHulls}

	\clearpage
	\bibliographystyle{chicago}
	\bibliography{./bibliography/Biblio}
	
\end{document}

%% file: images/Jumps.tex
	\begin{tikzpicture}
		\coordinate (left_end) at (-5, 0);
		\coordinate (right_end) at (5, 0);
		
		\pgfmathsetmacro{\step}{0.8}
		
		\draw (left_end) -- (right_end);
		\coordinate (p1) at (-4, 0);
		\coordinate (p2) at (-3, 0);
		\coordinate (p3) at (-1, 0);
		\coordinate (p4) at (1, 0);
		\coordinate (pn-1) at (3.5, 0);
		\coordinate (pn) at (4, 0);
		
		\pgfmathsetmacro{\vadjust}{-0.08}
		
		\node[very thin, circle, fill, inner sep = 2pt] (pp1) at (p1) {};
		\node[below] (p1_label) at (p1) {$\mathbf{L} = p_1$};
		\node[below] (p1_coordinate) at ($(p1)+(0, -0.5)$) {$x_1(t)$};
		\node[very thin, circle, fill, inner sep = 2pt] (pp2) at (p2) {};
		\node[below] (p2_label) at ($(p2) + (0, \vadjust)$) {$p_2$};
		\node[below] (pdots_label_1) at ($(p2)!0.5!(p3)$) {$\ldots$};
		\node[very thin, circle, fill, inner sep = 2pt] (pp3) at (p3) {};
		\node[below] (p3_label) at ($(p3) + (0, \vadjust)$) {$p_i$};
		\node[below] (pi_coordinate) at ($(p3)+(0, -0.5)$) {$x_i(t) = [x_i] + \{x_i\}$};
		\node[very thin, circle, fill, inner sep = 2pt] (pp4) at (p4) {};
		\node[below] (p4_label) at ($(p4) + (0, \vadjust)$) {$p_{i+1}$};
		\node[below] (pdots_label_2) at ($(p4)!0.5!(pn-1)$) {$\ldots$};
		\node[very thin, circle, fill, inner sep = 2pt] (ppn-1) at (pn-1) {};
		\node[below] (pn-1_label) at ($(pn-1) + (0, \vadjust)$) {$p_{n-1}$};
		\node[very thin, circle, fill, inner sep = 2pt] (ppn) at (pn) {};
		\node[below right] (pn_label) at (pn) {$p_N = \mathbf{R}$};
		\node[below] (pn_coordinate) at ($(pn)+(0, -0.5)$) {$x_N(t)$};
		
		\draw[densely dotted] ($(p1)+(-\step,0)$) circle (2pt);
		\draw[densely dotted] ($(p1)+(\step,0)$) circle (2pt);
		\draw[densely dotted] ($(pn)+(-\step,0)$) circle (2pt);
		\draw[densely dotted] ($(pn)+(\step,0)$) circle (2pt);
		
		\pgfmathsetmacro{\leftangle}{120}
		\pgfmathsetmacro{\rightangle}{60}
		\pgfmathsetmacro{\vertoffset}{0.3}
		
		\draw[-latex] (p1) to [out=\leftangle,in=\rightangle]($(p1)+(-\step,0)$);
		\draw[-latex] (p1) to [out=\rightangle,in=\leftangle]($(p1)+(\step,0)$);
		\node[above] (p1l_step) at ($(p1)+(-\step / 2, \vertoffset)$) {$-1$};
		\node[above] (p1r_step) at ($(p1)+(\step / 2, \vertoffset)$) {$+1$};
		\node[above] (p1l_probability) at ($(p1)+(-\step / 2, 3 * \vertoffset)$) {$\phantom{(}\varepsilon$};
		\node[above] (p1r_probability) at ($(p1)+(\step / 2, 3 * \vertoffset)$) {$(1-\varepsilon)$};
		\draw[-latex] (pn) to [out=\leftangle,in=\rightangle]($(pn)+(-\step,0)$);
		\draw[-latex] (pn) to [out=\rightangle,in=\leftangle]($(pn)+(\step,0)$);
		\node[above] (pnl_step) at ($(pn)+(-\step / 2, \vertoffset)$) {$-1$};
		\node[above] (pnr_step) at ($(pn)+(\step / 2, \vertoffset)$) {$+1$};
		\node[above] (pnl_probability) at ($(pn)+(\step / 2, 3 * \vertoffset)$) {$\phantom{(}\varepsilon$};
		\node[above] (pnr_probability) at ($(pn)+(-\step / 2, 3 * \vertoffset)$) {$(1-\varepsilon)$};
	\end{tikzpicture}

%% file: images/JumpSequence.tex
	\begin{tikzpicture}
		\coordinate (left_end) at (-5, 0);
		\coordinate (right_end) at (5, 0);
		
		\pgfmathsetmacro{\step}{0.8}
		
		\draw (left_end) -- (right_end);
		\coordinate (pn-1) at (-2, 0);
		\coordinate (pn) at (4, 0);
		
		\pgfmathsetmacro{\offset}{1.5}
		\node[very thin, circle, fill, inner sep = 1pt] (ppn0) at (pn) {};
		\node[very thin, circle, fill, inner sep = 1pt] (ppn1) at ($(pn)+(-\offset,0)$) {};
		\node[very thin, circle, fill, inner sep = 1pt] (ppn2) at ($(pn)+(-2*\offset,0)$) {};
		\node[below] (pn_label) at (pn) {$x_N(0)$};
		\node[very thin, circle, fill, inner sep = 1pt] (ppn-10) at (pn-1) {};
		\node[very thin, circle, fill, inner sep = 1pt] (ppn-11) at ($(pn-1)+(-\offset,0)$) {};
		\pgfmathsetmacro{\hoffset}{0.5}
		\node[below] (pn_1_label) at ($(pn-1)+(-\hoffset, 0)$) {$x_{N-1}(0)$};
		
		\pgfmathsetmacro{\height}{0.2}
		\draw[thick] ($(pn)+(0, \height)$) -- ($(pn)+(0, -\height)$);
		\draw[thick] ($(pn-1)+(-\hoffset, \height)$) -- ($(pn-1)+(-\hoffset, -\height)$);
		
		\pgfmathsetmacro{\leftangle}{120}
		\pgfmathsetmacro{\rightangle}{60}
		\draw[-latex] (ppn0) to [out=\leftangle,in=\rightangle](ppn1);
		\draw[-latex] (ppn1) to [out=\leftangle,in=\rightangle](ppn2);
		\draw[-latex] (ppn-10) to [out=\leftangle,in=\rightangle](ppn-11);
		\node[above] (pnl_dots) at ($(ppn2)!.5!(ppn-10)+(0, 0.2)$) {$\cdots$};
		\node[above] (pnl_value1) at ($(ppn0)!.5!(ppn1)+(0, 0.4)$) {-1};
		\node[above] (pnl_value2) at ($(ppn1)!.5!(ppn2)+(0, 0.4)$) {-1};
		\node[above] (pnl_value3) at ($(ppn-10)!.5!(ppn-11)+(0, 0.4)$) {-1};
	\end{tikzpicture}

%% file: images/InternalAgentsGatheredNearJump.tex
	\begin{tikzpicture}
		\coordinate (left_end_T) at (-1, 0);
		\coordinate (right_end_T) at (8, 0);
		
		\coordinate (x1_T) at (0, 0);
		\coordinate (x2_T) at (1, 0);
		\coordinate (xN1_T) at (6, 0);
		\coordinate (x1_T1) at (5, 0);
		
		\draw[thick] (left_end_T) -- (right_end_T);
		
		\pgfmathsetmacro{\ysmalloffset}{-0.1}
		\draw[color=white,pattern=north west lines, pattern color=blue] ($(x2_T)+(0,\ysmalloffset)$) rectangle ($(xN1_T)+(0,-\ysmalloffset)$);
		
		\node[very thin, circle, fill, inner sep = 1pt] (px1_T) at (x1_T) {};
		\node[very thin, circle, fill, inner sep = 1pt] (px2_T) at (x2_T) {};
		\node[very thin, circle, fill, inner sep = 1pt] (pxN1_T) at (xN1_T) {};
		\node[below right] (px2_T_label) at (x2_T) {$x_{2}(T)$};
		\node[below] (pxN1_T_label) at (xN1_T) {$x_{N-1}(T)$};
		\node[below left] (px1_T_label) at (x1_T) {$x_{1}(T)$};
		\node[above] (A_L_label) at ($(x2_T) + (0, 0.5)$) {$A_L$};
		\node[above] (A_R_label) at ($(xN1_T) + (0, 0.5)$) {$A_R$};

		\path[-latex, bend left, thick] (x1_T) edge (x1_T1);
	\end{tikzpicture}

%% file: images/InternalAgentsGatheredFarJump.tex
	\begin{tikzpicture}
		\coordinate (left_end_T) at (-1, 0);
		\coordinate (right_end_T) at (8, 0);
		
		\coordinate (x1_T) at (0, 0);
		\coordinate (x2_T) at (1, 0);
		\coordinate (xN1_T) at (6, 0);
		\coordinate (x1_T1) at (7, 0);
		
		\draw[thick] (left_end_T) -- (right_end_T);
		
		\pgfmathsetmacro{\ysmalloffset}{-0.1}
		\draw[color=white,pattern=north west lines, pattern color=blue] ($(x2_T)+(0,\ysmalloffset)$) rectangle ($(xN1_T)+(0,-\ysmalloffset)$);
		\draw[color=white,pattern=north east lines, pattern color=blue] ($(xN1_T)+(0,\ysmalloffset)$) rectangle ($(x1_T1)+(0,-\ysmalloffset)$);
		
		\node[very thin, circle, fill, inner sep = 1pt] (px1_T) at (x1_T) {};
		\node[very thin, circle, fill, inner sep = 1pt] (px2_T) at (x2_T) {};
		\node[very thin, circle, fill, inner sep = 1pt] (pxN1_T) at (xN1_T) {};
		\node[below right] (px2_T_label) at (x2_T) {$x_{2}(T)$};
		\node[below] (pxN1_T_label) at (xN1_T) {$x_{N-1}(T)$};
		\node[below left] (px1_T_label) at (x1_T) {$x_{1}(T)$};
		\node[above] (A_L_label) at ($(x2_T) + (0, 0.5)$) {$A_L$};
		\node[above] (A_R_label) at ($(xN1_T) + (0, 0.5)$) {$A_R$};

		\path[-latex, bend left, thick] (x1_T) edge (x1_T1);
	\end{tikzpicture}

%% file: images/RPMarkovChain.tex
	\begin{tikzpicture}[->, >={Stealth}, auto, semithick, node distance=1.5cm]
		\tikzstyle{every state}=[fill=white,draw=black,thick,text=black]
		\node[state]    		(1)					{$1$};
		\node[state]    		(2)[right=of 1]		{$2$};
		\node[state]    		(3)[right=of 2]		{$3$};
		\node[state]    		(4)[right=of 3]		{$4$};
		\node[state, draw=none]	(5)[right=of 4]		{\ldots};
		\path
		(1) edge[bend left]		node{$\varepsilon$}				(2)
			edge[loop left]     node{$1-\varepsilon$}			(1)
		(2)edge[bend left]     node{$\varepsilon$}				(3)
			edge[bend left]		node{$1-\varepsilon$}			(1)
		(3)edge[bend left]     node{$\varepsilon$}				(4)
			edge[bend left]		node{$1-\varepsilon$}			(2)
		(4)edge[bend left]     node{$\varepsilon$}				(5)
			edge[bend left]		node{$1-\varepsilon$}			(3)
		(5)edge[bend left]		node{$1-\varepsilon$}			(4);
	\end{tikzpicture}

%% file: images/ConvergenceTimeN400S0500.Theoretical.tex
%
%
\begin{tikzpicture}

\begin{axis}[%
width=4.521in,
height=3.566in,
at={(0.758in,0.481in)},
scale only axis,
xmin=0,
xmax=0.5,
xlabel style={font=\color{white!15!black}\large},
xlabel={$\varepsilon$},
ymin=0,
ymax=1254000,
ylabel style={font=\color{white!15!black}\large},
ylabel={$\expectation{T}$ (\# of steps)},
axis background/.style={fill=white},
title style={font=\bfseries},
legend style={at={(0.03,0.97)}, anchor=north west, legend cell align=left, align=left, draw=white!15!black}
]
\addplot [color=blue, mark=asterisk, mark options={solid, blue}]
  table[row sep=crcr]{%
0.0100000000093132	25741.02\\
0.0200000000186265	26221.53\\
0.0300000000279397	26901.8799999999\\
0.0400000000372529	27503.51\\
0.0500000000465661	28010.8999999999\\
0.0600000000558794	28568.1799999999\\
0.0700000000651926	29239.76\\
0.0800000000745058	30106.1799999999\\
0.090000000083819	30779.1100000001\\
0.100000000093132	31443.0800000001\\
0.110000000102445	32310.6100000001\\
0.120000000111759	33104.21\\
0.129999999888241	33945.8999999999\\
0.139999999897555	34769.9099999999\\
0.149999999906868	35764.05\\
0.159999999916181	36983.3799999999\\
0.169999999925494	38251.8300000001\\
0.179999999934807	39364.95\\
0.189999999944121	40436.49\\
0.199999999953434	42171.3799999999\\
0.209999999962747	43428.49\\
0.21999999997206	45056.5\\
0.229999999981374	46762.45\\
0.239999999990687	48550.1599999999\\
0.25	50509.8\\
0.260000000009313	52808.1699999999\\
0.270000000018626	54953.29\\
0.28000000002794	57389.3899999999\\
0.290000000037253	60128.5800000001\\
0.300000000046566	62951.3899999999\\
0.310000000055879	66414.26\\
0.320000000065193	69776.45\\
0.330000000074506	73926\\
0.340000000083819	78885.28\\
0.350000000093132	83972.1399999999\\
0.360000000102445	89863.5800000001\\
0.370000000111759	96623.3400000001\\
0.379999999888241	104995.17\\
0.389999999897555	115411.17\\
0.399999999906868	126065.35\\
0.409999999916181	139949.52\\
0.419999999925494	157939.69\\
0.429999999934807	180620.05\\
0.439999999944121	210478.89\\
0.449999999953434	251848.68\\
0.459999999962747	314107.89\\
0.46999999997206	421185.04\\
0.479999999981374	630362.02\\
0.489999999990687	1253315.9\\
};
\addlegendentry{Actual averages of 100 runs}

\addplot [color=red, mark=*, mark options={solid, red}]
  table[row sep=crcr]{%
0.0100000000093132	204492.35333231\\
0.0200000000186265	208752.6106934\\
0.0300000000279397	213194.15560177\\
0.0400000000372529	217828.811158331\\
0.0500000000465661	222669.451406294\\
0.0600000000558794	227730.120756437\\
0.0700000000651926	233026.170076354\\
0.0800000000745058	238574.412221029\\
0.090000000083819	244393.300323981\\
0.100000000093132	250503.13283208\\
0.110000000102445	256926.290084185\\
0.120000000111759	263687.508244295\\
0.129999999888241	270814.197656303\\
0.139999999897555	278336.814257867\\
0.149999999906868	286289.294665235\\
0.159999999916181	294709.568037741\\
0.169999999925494	303640.161008582\\
0.179999999934807	313128.9160401\\
0.189999999944121	323229.848815587\\
0.199999999953434	334004.17710944\\
0.209999999962747	345521.562527007\\
0.21999999997206	357861.618331543\\
0.229999999981374	371115.752343823\\
0.239999999990687	385389.435126277\\
0.25	400805.012531328\\
0.260000000009313	417505.2213868\\
0.270000000018626	435657.622316661\\
0.28000000002794	455460.241512873\\
0.290000000037253	477148.824442057\\
0.300000000046566	501006.26566416\\
0.310000000055879	527375.01648859\\
0.320000000065193	556673.628515734\\
0.330000000074506	589419.136075483\\
0.340000000083819	626257.832080201\\
0.350000000093132	668008.354218881\\
0.360000000102445	715723.236663086\\
0.370000000111759	770778.870252555\\
0.379999999888241	835010.442773601\\
0.389999999897555	910920.483025746\\
0.399999999906868	1002012.53132832\\
0.409999999916181	1113347.25703147\\
0.419999999925494	1252515.6641604\\
0.427091251360253	1379400\\
};
\addlegendentry{Theoretical upper bound}

\end{axis}
\end{tikzpicture}%

%% file: images/PredictionRateN400S0500.tex
%
%
\begin{tikzpicture}

\begin{axis}[%
width=4.521in,
height=3.566in,
at={(0.758in,0.481in)},
scale only axis,
xmin=0,
xmax=0.5,
xlabel style={font=\color{white!15!black}\large},
xlabel={$\varepsilon$},
ymin=7.88,
ymax=8.02,
ylabel style={font=\color{white!15!black}\large},
ylabel={$\expectation{T}$ bound / Actual $\expectation{T}$},
axis background/.style={fill=white},
title style={font=\bfseries},
legend style={at={(0.03,0.97)}, anchor=north west, legend cell align=left, align=left, draw=white!15!black}
]
\addplot [color=blue, mark=asterisk, mark options={solid, blue}]
  table[row sep=crcr]{%
0.00999999999999979	7.94422106553316\\
0.0199999999999996	7.96111480502473\\
0.0299999999999994	7.92487943600114\\
0.0399999999999991	7.92003679378852\\
0.0500000000000007	7.9493858250286\\
0.0600000000000005	7.9714605815434\\
0.0700000000000003	7.9694966742666\\
0.0800000000000001	7.9244331968064\\
0.0899999999999999	7.94023285026697\\
0.0999999999999996	7.96687642661216\\
0.109999999999999	7.95176228750199\\
0.119999999999999	7.96537685823933\\
0.130000000000001	7.97781757609322\\
0.140000000000001	8.00510597404097\\
0.15	8.0049461586491\\
0.16	7.96870291568108\\
0.17	7.93792508772998\\
0.18	7.95451070152764\\
0.19	7.99351894330065\\
0.199999999999999	7.92016237337835\\
0.210000000000001	7.95610352851336\\
0.220000000000001	7.94250814713844\\
0.23	7.93619137457132\\
0.24	7.93796426471668\\
0.25	7.93519302256846\\
0.26	7.90607251466582\\
0.27	7.92778052627352\\
0.279999999999999	7.93631438690798\\
0.289999999999999	7.93547468511742\\
0.300000000000001	7.95862117840703\\
0.310000000000001	7.9406894918138\\
0.32	7.9779585879725\\
0.33	7.97309655703653\\
0.34	7.93884273568149\\
0.35	7.95511885512124\\
0.359999999999999	7.96455289966287\\
0.369999999999999	7.97714993346902\\
0.380000000000001	7.95284623829459\\
0.390000000000001	7.89282773084916\\
0.4	7.94835798519039\\
0.41	7.95534887887767\\
0.42	7.93034141171482\\
0.43	7.92518036245795\\
0.44	7.93438660545578\\
0.449999999999999	7.95725855166937\\
0.460000000000001	7.97506655538262\\
0.470000000000001	7.93010542609587\\
0.48	7.9479132588629\\
0.49	7.99489204061259\\
};
\addlegendentry{Bound to actual rate}

\end{axis}
\end{tikzpicture}%

%% file: images/ConvergenceTimeEpsilon0.1N400.Theoretical.tex
%
%
\begin{tikzpicture}

\begin{axis}[%
width=4.521in,
height=3.517in,
at={(0.758in,0.53in)},
scale only axis,
xmin=0,
xmax=1000,
xlabel style={font=\color{white!15!black}},
xlabel={$S_0$},
ymin=0,
ymax=600000,
ylabel style={font=\color{white!15!black}},
ylabel={$\expectation{T}$ (\# of steps)},
axis background/.style={fill=white},
legend style={at={(0.03,0.97)}, anchor=north west, legend cell align=left, align=left, draw=white!15!black}
]
\addplot [color=blue, mark size=1.5pt, mark=asterisk, mark options={solid, blue}]
  table[row sep=crcr]{%
50	3142.87\\
100	6328.46\\
150	9452.07\\
200	12622.97\\
250	15718.64\\
300	18928.4\\
350	22086.84\\
400	25066.34\\
450	28352.9\\
500	31543.91\\
550	34705.73\\
600	37859.02\\
650	41097.52\\
700	44140.51\\
750	47267.81\\
800	50410.56\\
850	53484.24\\
900	56649.55\\
950	59894.78\\
1000	63098.45\\
};
\addlegendentry{Actual averages of 100 runs}

\addplot [color=red, mark size=2.0pt, mark=*, mark options={solid, red}]
  table[row sep=crcr]{%
50	22000.313283208\\
100	48000.6265664161\\
150	74000.9398496241\\
200	99001.2531328321\\
250	124501.56641604\\
300	150001.879699248\\
350	175002.192982456\\
400	200002.506265664\\
450	225502.819548872\\
500	250503.13283208\\
550	275503.446115288\\
600	301003.759398496\\
650	326004.072681704\\
700	351004.385964912\\
750	376004.69924812\\
800	401005.012531328\\
850	426505.325814536\\
900	451505.639097744\\
950	476505.952380952\\
1000	501506.26566416\\
};
\addlegendentry{Theoretical upper bound}

\end{axis}
\end{tikzpicture}%

%% file: images/PredictionRateEpsilon0.1N400.tex
%
%
\begin{tikzpicture}

\begin{axis}[%
width=4.521in,
height=3.517in,
at={(0.758in,0.53in)},
scale only axis,
xmin=0,
xmax=1000,
xlabel style={font=\color{white!15!black}},
xlabel={$S_0$},
ymin=7,
ymax=8,
ylabel style={font=\color{white!15!black}},
ylabel={$\expectation{T}$ bound / Actual $\expectation{T}$},
axis background/.style={fill=white},
title style={font=\bfseries},
legend style={at={(0.97,0.03)}, anchor=south east, legend cell align=left, align=left, draw=white!15!black}
]
\addplot [color=blue, mark=asterisk, mark options={solid, blue}]
  table[row sep=crcr]{%
50	7.00007104436645\\
100	7.58488266757092\\
150	7.82907234601771\\
200	7.84294449981519\\
250	7.92063221856597\\
300	7.92469937761507\\
350	7.92336943548537\\
400	7.97892736896029\\
450	7.95343049736971\\
500	7.94141033347103\\
550	7.93826973572629\\
600	7.95064846893808\\
650	7.93245122045573\\
700	7.95197848789951\\
750	7.95477301038738\\
800	7.95478194511884\\
850	7.97441126235572\\
900	7.97015402766203\\
950	7.95571754969217\\
1000	7.9479965936431\\
};
\addlegendentry{Bound to actual rate}

\end{axis}
\end{tikzpicture}%

%% file: images/ConvergenceTimeEpsilon0.1S01000.Theoretical.tex
%
%
\begin{tikzpicture}

\begin{axis}[%
width=4.521in,
height=3.566in,
at={(0.758in,0.481in)},
scale only axis,
xmin=0,
xmax=1000,
xlabel style={font=\color{white!15!black}},
xlabel={N},
ymin=0,
ymax=1400000,
ylabel style={font=\color{white!15!black}},
ylabel={$\expectation{T}$ (\# of steps)},
axis background/.style={fill=white},
legend style={at={(0.03,0.97)}, anchor=north west, legend cell align=left, align=left, draw=white!15!black}
]
\addplot [color=blue, mark size=2.0pt, mark=asterisk, mark options={solid, blue}]
  table[row sep=crcr]{%
50	8402.38\\
100	16227.46\\
150	23847.48\\
200	31787.65\\
250	39705.28\\
300	47628.9\\
350	55136.69\\
400	62963.44\\
450	70566.7\\
500	78526.1\\
550	86261.36\\
600	94257.88\\
650	102207.51\\
700	110257.85\\
750	117513.7\\
800	125522.03\\
850	133755.97\\
900	141338.44\\
950	148947.77\\
1000	156615.12\\
};
\addlegendentry{Actual averages of 100 runs}

\addplot [color=red, mark size=2.0pt, mark=*, mark options={solid, red}]
  table[row sep=crcr]{%
50	63113.5204081633\\
100	125900.252525253\\
150	188641.77852349\\
200	251262.56281407\\
250	313760.040160642\\
300	376508.361204013\\
350	439257.163323782\\
400	501506.26566416\\
450	564193.067928731\\
500	626255.01002004\\
550	688879.553734062\\
600	751504.173622705\\
650	814128.852080123\\
700	876753.576537911\\
750	938440.837783712\\
800	1001003.12891114\\
850	1063565.44464075\\
900	1126127.78086763\\
950	1188690.13435195\\
1000	1250002.5025025\\
};
\addlegendentry{Theoretical upper bound}

\end{axis}
\end{tikzpicture}%

%% file: images/PredictionRateEpsilon0.1S01000.tex
%
%
\begin{tikzpicture}

\begin{axis}[%
width=4.521in,
height=3.566in,
at={(0.758in,0.481in)},
scale only axis,
xmin=0,
xmax=1000,
xlabel style={font=\color{white!15!black}},
xlabel={N},
ymin=7.5,
ymax=8,
ylabel style={font=\color{white!15!black}},
ylabel={$\expectation{T}$ bound / Actual $\expectation{T}$},
axis background/.style={fill=white},
title style={font=\bfseries},
legend style={at={(0.97,0.03)}, anchor=south east, legend cell align=left, align=left, draw=white!15!black}
]
\addplot [color=blue, mark=asterisk, mark options={solid, blue}]
  table[row sep=crcr]{%
50	7.51138610824114\\
100	7.75846944162868\\
150	7.91034434344806\\
200	7.9044082470416\\
250	7.90222459483073\\
300	7.90504003250157\\
350	7.96669447012107\\
400	7.96503916660458\\
450	7.9951743234235\\
500	7.97511922812976\\
550	7.98595748703781\\
600	7.97285249384674\\
650	7.965450406532\\
700	7.95184720668783\\
750	7.98579942409879\\
800	7.97472068378067\\
850	7.9515362539762\\
900	7.96759735615899\\
950	7.98058362573636\\
1000	7.98136541671397\\
};
\addlegendentry{Bound to actual rate}

\end{axis}
\end{tikzpicture}%

%% file: images/N.TotalDistance.semilogy.tex
%
%
\definecolor{mycolor1}{rgb}{0.00000,1.00000,1.00000}%
\definecolor{mycolor2}{rgb}{1.00000,0.00000,1.00000}%
\begin{tikzpicture}

\begin{axis}[%
width=4.521in,
height=3.566in,
at={(0.758in,0.481in)},
scale only axis,
xmin=0,
xmax=20,
xlabel style={font=\color{white!15!black}},
xlabel={Total Distance, $d$},
ymode=log,
ymin=1e-06,
ymax=100,
yminorticks=true,
ylabel style={font=\color{white!15!black}},
ylabel={\% of time Span $> d$},
axis background/.style={fill=white},
title style={font=\bfseries},
legend style={legend cell align=left, align=left, draw=white!15!black}
]
\addplot [color=blue, mark=*, mark options={solid, blue}]
  table[row sep=crcr]{%
1	82.2332726454236\\
2	14.7304919792488\\
3	2.6532301522674\\
4	0.327329761774316\\
5	0.0493848939593326\\
6	0.00541053453807902\\
7	0.000785057952584014\\
8	8.68918711997751e-05\\
9	8.08296476276977e-06\\
};
\addlegendentry{N = 10}

\addplot [color=green, mark=o, mark options={solid, green}]
  table[row sep=crcr]{%
1	80.7743387257168\\
2	16.0083188551602\\
3	2.7788922845516\\
4	0.376416540373635\\
5	0.0544075150898556\\
6	0.00666326751723107\\
7	0.000852775980416268\\
8	9.16963419802441e-05\\
9	1.73204201518239e-05\\
10	1.01884824422493e-06\\
};
\addlegendentry{N = 20}

\addplot [color=red, mark=x, mark options={solid, red}]
  table[row sep=crcr]{%
1	79.7527209073316\\
2	16.909472066622\\
3	2.8628922604282\\
4	0.408895242910858\\
5	0.0576443162088041\\
6	0.00730966920978723\\
7	0.000922701076973568\\
8	0.000125112010437094\\
9	1.35538011306852e-05\\
10	4.17040034790313e-06\\
};
\addlegendentry{N = 50}

\addplot [color=mycolor1, mark=+, mark options={solid, mycolor1}]
  table[row sep=crcr]{%
1	79.393160825617\\
2	17.222087065128\\
3	2.895425983985\\
4	0.420771756642845\\
5	0.0595352911897335\\
6	0.00789455587042523\\
7	0.000969640808069636\\
8	0.000129794438875463\\
9	2.5086320118787e-05\\
};
\addlegendentry{N = 100}

\addplot [color=mycolor2, mark=asterisk, mark options={solid, mycolor2}]
  table[row sep=crcr]{%
1	79.2059570604625\\
2	17.3862325894588\\
3	2.91104223442338\\
4	0.427601360526896\\
5	0.0602229221728679\\
6	0.00776261702374844\\
7	0.00103207250623154\\
8	0.000124087330228995\\
9	2.38629481209605e-05\\
10	1.19314740604802e-06\\
};
\addlegendentry{N = 200}

\end{axis}
\end{tikzpicture}%

%% file: images/N.WBounds.TotalDistanceCDF.semilogy.tex
%
%
\definecolor{mycolor1}{rgb}{0.00000,1.00000,1.00000}%
\definecolor{mycolor2}{rgb}{1.00000,0.00000,1.00000}%
\definecolor{mycolor3}{rgb}{0.5,0.5,0.5}%
\begin{tikzpicture}

\begin{axis}[%
width=4.521in,
height=3.566in,
at={(0.758in,0.481in)},
scale only axis,
xmin=0,
xmax=20,
xlabel style={font=\color{white!15!black}\large},
xlabel={Total Distance, $d$},
ymode=log,
ymin=0.5,
ymax=1,
yminorticks=true,
ylabel style={font=\color{white!15!black}\large},
ylabel={$\probability{\text{Span} < d}$},
axis background/.style={fill=white},
title style={font=\bfseries},
legend style={at={(0.97,0.03)}, anchor=south east, legend cell align=left, align=left, draw=white!15!black}
]
\addplot [color=blue, mark=*, mark options={solid, blue}]
  table[row sep=crcr]{%
1	0.434884059962229\\
2	0.965682096953906\\
3	0.990656253745173\\
4	0.999303195686726\\
5	0.999853720928769\\
6	0.999988223680604\\
7	0.999997907555268\\
8	0.999999858482002\\
9	0.999999949457856\\
10	1\\
11	1\\
12	1\\
13	1\\
14	1\\
15	1\\
16	1\\
17	1\\
18	1\\
19	1\\
};
\addlegendentry{N = 10}

\addplot [color=green, mark=o, mark options={solid, green}]
  table[row sep=crcr]{%
1	0.43487726369148\\
2	0.965713167732389\\
3	0.990655959466439\\
4	0.99930488957808\\
5	0.999854313784519\\
6	0.999988054667221\\
7	0.999997688328608\\
8	0.999999786144941\\
9	0.999999949082129\\
10	1\\
11	1\\
12	1\\
13	1\\
14	1\\
15	1\\
16	1\\
17	1\\
18	1\\
19	1\\
};
\addlegendentry{N = 20}

\addplot [color=red, mark=x, mark options={solid, red}]
  table[row sep=crcr]{%
1	0.434804690594699\\
2	0.965653434308323\\
3	0.990652133481625\\
4	0.999298801413107\\
5	0.999853611964237\\
6	0.999987981123532\\
7	0.999997577428895\\
8	0.999999853810366\\
9	0.999999979115766\\
10	0.999999989557883\\
11	1\\
12	1\\
13	1\\
14	1\\
15	1\\
16	1\\
17	1\\
18	1\\
19	1\\
};
\addlegendentry{N = 50}

\addplot [color=mycolor1, mark=+, mark options={solid, mycolor1}]
  table[row sep=crcr]{%
1	0.434777362906234\\
2	0.965680564586753\\
3	0.990647776653816\\
4	0.999304654232752\\
5	0.999856633998081\\
6	0.9999890078092\\
7	0.999997945080054\\
8	0.999999847783708\\
9	0.999999945637041\\
10	1\\
11	1\\
12	1\\
13	1\\
14	1\\
15	1\\
16	1\\
17	1\\
18	1\\
19	1\\
};
\addlegendentry{N = 100}

\addplot [color=mycolor2, mark=asterisk, mark options={solid, mycolor2}]
  table[row sep=crcr]{%
1	0.434604011307683\\
2	0.96561414957107\\
3	0.990654342775789\\
4	0.999304640939476\\
5	0.999854075739988\\
6	0.999987896232493\\
7	0.999997796065952\\
8	0.999999797476333\\
9	0.999999928521056\\
10	1\\
11	1\\
12	1\\
13	1\\
14	1\\
15	1\\
16	1\\
17	1\\
18	1\\
19	1\\
};
\addlegendentry{N = 200}

\addplot [dashed, color=black, mark=diamond, mark options={solid, black}]
  table[row sep=crcr]{%
3	0.790123456790123\\
4	0.965706447187928\\
5	0.994970278920899\\
6	0.999305661399851\\
7	0.999907797855261\\
8	0.999988082715987\\
9	0.999998490012749\\
10	0.999999811574239\\
11	0.999999976769424\\
12	0.999999997163896\\
13	0.999999999656553\\
14	0.999999999958689\\
15	0.999999999995059\\
16	0.999999999999411\\
17	0.999999999999926\\
18	0.999999999999992\\
19	1\\
};
\addlegendentry{Theoretical lower bound (\autoref{thm:IntervalDistributions})}

\addplot [dotted, color=black, mark=*, mark options={solid, black}]
  table[row sep=crcr]{%
2	0.5\\
3	0.666666666666667\\
4	0.75\\
5	0.8\\
6	0.833333333333333\\
7	0.857142857142857\\
8	0.875\\
9	0.888888888888889\\
10	0.900000000000001\\
11	0.909090909090908\\
12	0.916666666666667\\
13	0.923076923076923\\
14	0.928571428571429\\
15	0.933333333333333\\
16	0.9375\\
17	0.941176470588232\\
18	0.944444444444448\\
19	0.947368421052633\\
};
\addlegendentry{Theoretical lower bound (crude) (Equation \eqref{eq:EasyBound})}

\end{axis}
\end{tikzpicture}%

%% file: images/centers.3.tex
\begin{tikzpicture}

\begin{axis}[%
width=7.083in,
height=5in,
at={(0.625in,0.625in)},
scale only axis,
xmin=-100,
xmax=0,
x tick label style={rotate=90,anchor=east},
xticklabel={\round[0]{10100+\tick}},
xlabel style={font=\color{white!15!black}},
xlabel={Iteration},
ymin=-8,
ymax=4,
ylabel style={font=\color{white!15!black}},
ylabel={Offset},
axis background/.style={fill=white},
title style={font=\bfseries},
axis x line*=bottom,
axis y line*=left,
legend style={at={(0.03,0.97)}, anchor=north west, legend cell align=left, align=left, draw=white!15!black}
]
\addplot[const plot, color=blue, line width=2pt] table[row sep=crcr] {%
-99	-3.29913937979148\\
-98	-3.34600759163931\\
-97	-3.39363604909573\\
-96	-3.34594547871188\\
-95	-3.29913937979148\\
-90	-3.24821332950269\\
-89	-3.20114489344431\\
-86	-3.15072718959665\\
-83	-3.10999366897057\\
-80	-3.16041137281825\\
-79	-3.15881821968024\\
-78	-3.10777979387088\\
-77	-3.0603448465707\\
-76	-3.01934967898931\\
-74	-2.96937480050956\\
-73	-2.91790397764247\\
-72	-2.97051160760331\\
-71	-3.01934967898931\\
-69	-3.06415187480842\\
-67	-3.01934967898931\\
-66	-3.06678462628949\\
-65	-3.10999366897057\\
-61	-3.15881821968024\\
-59	-3.10999366897057\\
-57	-3.15881821968024\\
-56	-3.20367278063891\\
-54	-3.15881821968024\\
-53	-3.10777979387088\\
-52	-3.0603448465707\\
-51	-3.01034601910442\\
-50	-2.95773838914357\\
-43	-2.91790397764247\\
-42	-2.86780264212631\\
-41	-2.81559426495282\\
-40	-2.76328881808067\\
-37	-2.71469869239945\\
-35	-2.71911507677351\\
-34	-2.77142052364566\\
-33	-2.7673790703795\\
-32	-2.71878894469829\\
-29	-2.66923527697013\\
-28	-2.61946218767841\\
-23	-2.5698523587067\\
-22	-2.58015634492293\\
-17	-2.62690768423899\\
-16	-2.67646135196713\\
-15	-2.72505147764835\\
-14	-2.7773569245205\\
-13	-2.829565301694\\
-12	-2.87966663721016\\
-11	-2.93116140906378\\
-6	-2.98376903902461\\
-3	-3.0337678664909\\
-1	-3.01934967898931\\
0	-3.29913937979148\\
};
\addlegendentry{Core-Center}

\addplot[const plot, color=red, dashdotted, line width=2pt] table[row sep=crcr] {%
-99	-3.78382771499962\\
-98	-4.78382771499962\\
-97	-3.87770687770634\\
-96	-3.78382771499962\\
-95	-3.76820290281509\\
-94	-3.78382771499962\\
-93	-3.76820290281509\\
-92	-4.76820290281509\\
-91	-3.76820290281509\\
-90	-3.73579785830229\\
-89	-3.63009814341135\\
-88	-4.63009814341135\\
-87	-3.63009814341135\\
-86	-3.58803451651703\\
-85	-4.58803451651703\\
-84	-3.58803451651703\\
-83	-3.55776460689485\\
-82	-3.58803451651703\\
-81	-3.55776460689485\\
-80	-4.55776460689485\\
-79	-3.58803451651703\\
-78	-3.55776460689485\\
-77	-3.45902860559818\\
-76	-3.40900632745776\\
-75	-3.45902860559818\\
-74	-3.4085512967137\\
-73	-3.38695196193228\\
-72	-4.38695196193228\\
-71	-3.40900632745776\\
-70	-3.45902860559818\\
-69	-4.45902860559818\\
-68	-3.45902860559818\\
-67	-3.40900632745776\\
-66	-4.40900632745776\\
-65	-3.55776460689485\\
-64	-3.58803451651703\\
-63	-3.55776460689485\\
-62	-3.58803451651703\\
-61	-4.58803451651703\\
-60	-3.58803451651703\\
-59	-3.55776460689485\\
-58	-3.58803451651703\\
-57	-4.58803451651703\\
-56	-3.63009814341135\\
-55	-3.73579785830229\\
-54	-3.58803451651703\\
-53	-3.55776460689485\\
-52	-3.45902860559818\\
-51	-3.40900632745776\\
-50	-3.4085512967137\\
-49	-4.4085512967137\\
-48	-3.4085512967137\\
-47	-4.4085512967137\\
-46	-3.4085512967137\\
-45	-4.4085512967137\\
-44	-3.4085512967137\\
-43	-3.38695196193228\\
-42	-3.3388773367393\\
-41	-3.33083650303571\\
-40	-3.32463999360667\\
-39	-4.32463999360667\\
-38	-3.32463999360667\\
-37	-3.24785238154978\\
-36	-4.24785238154978\\
-35	-3.24785238154978\\
-34	-4.24785238154978\\
-33	-3.32463999360667\\
-32	-3.24785238154978\\
-31	-4.24785238154978\\
-30	-3.24785238154978\\
-29	-3.18937206838477\\
-28	-3.13506076492754\\
-27	-4.13506076492754\\
-26	-3.13506076492754\\
-25	-4.13506076492754\\
-24	-3.13506076492754\\
-23	-3.07764751538988\\
-22	-4.07764751538988\\
-21	-5.07764751538988\\
-20	-4.07764751538988\\
-19	-3.07764751538988\\
-18	-4.07764751538988\\
-17	-3.13506076492754\\
-16	-4.13506076492754\\
-15	-5.13506076492754\\
-14	-6.13506076492754\\
-13	-5.13506076492754\\
-12	-4.13506076492754\\
-11	-5.13506076492754\\
-10	-4.13506076492754\\
-9	-5.13506076492754\\
-8	-6.13506076492754\\
-7	-7.13506076492754\\
-6	-6.13506076492754\\
-5	-5.13506076492754\\
-4	-6.13506076492754\\
-3	-5.13506076492754\\
-2	-4.13506076492754\\
-1	-3.40900632745776\\
0	-3.76820290281509\\
};
\addlegendentry{Left Extremal Point}

\addplot[const plot, color=purple, dashed, line width=2pt] table[row sep=crcr] {%
-99	-2.76820290281509\\
-98	-2.87770687770634\\
-97	-2.87888702332754\\
-96	-1.87888702332754\\
-95	-2.78382771499962\\
-94	-2.76820290281509\\
-93	-2.78382771499962\\
-92	-1.78382771499962\\
-91	-0.783827714999617\\
-90	-1.78382771499962\\
-89	-0.783827714999617\\
-88	-1.78382771499962\\
-87	-0.783827714999617\\
-86	0.216172285000383\\
-85	-0.783827714999617\\
-84	-1.78382771499962\\
-83	-2.58803451651703\\
-82	-2.55776460689485\\
-81	-2.58803451651703\\
-80	-2.63009814341135\\
-79	-1.63009814341135\\
-78	-0.630098143411345\\
-77	-1.63009814341135\\
-76	-2.45902860559818\\
-75	-2.40900632745776\\
-74	-1.40900632745776\\
-73	-2.4085512967137\\
-72	-2.40900632745776\\
-71	-2.45902860559818\\
-70	-2.40900632745776\\
-69	-2.55776460689485\\
-68	-1.55776460689485\\
-67	-2.45902860559818\\
-66	-2.55776460689485\\
-65	-2.58803451651703\\
-64	-2.55776460689485\\
-63	-2.58803451651703\\
-62	-2.55776460689485\\
-61	-2.63009814341135\\
-60	-1.63009814341135\\
-59	-2.58803451651703\\
-58	-2.55776460689485\\
-57	-2.63009814341135\\
-56	-2.73579785830229\\
-55	-2.63009814341135\\
-54	-1.63009814341135\\
-53	-0.630098143411345\\
-52	0.369901856588655\\
-51	-0.630098143411345\\
-50	0.369901856588655\\
-49	1.36990185658865\\
-48	2.36990185658865\\
-47	1.36990185658865\\
-46	0.369901856588655\\
-45	-0.630098143411345\\
-44	-1.63009814341135\\
-43	-2.4085512967137\\
-42	-1.4085512967137\\
-41	-0.408551296713696\\
-40	0.591448703286304\\
-39	1.5914487032863\\
-38	0.591448703286304\\
-37	-0.408551296713696\\
-36	-1.4085512967137\\
-35	-2.32463999360667\\
-34	-2.33083650303571\\
-33	-1.33083650303571\\
-32	-0.330836503035712\\
-31	0.669163496964288\\
-30	1.66916349696429\\
-29	2.66916349696429\\
-28	1.66916349696429\\
-27	0.669163496964288\\
-26	1.66916349696429\\
-25	0.669163496964288\\
-24	-0.330836503035712\\
-23	-1.33083650303571\\
-22	-2.13506076492754\\
-21	-1.13506076492754\\
-20	-2.13506076492754\\
-19	-1.13506076492754\\
-18	-2.13506076492754\\
-17	-2.18937206838477\\
-16	-2.24785238154978\\
-15	-2.32463999360667\\
-14	-2.33083650303571\\
-13	-2.3388773367393\\
-12	-2.38695196193228\\
-11	-2.4085512967137\\
-10	-1.4085512967137\\
-9	-0.408551296713696\\
-8	-1.4085512967137\\
-7	-2.4085512967137\\
-6	-2.40900632745776\\
-5	-1.40900632745776\\
-4	-2.40900632745776\\
-3	-2.45902860559818\\
-2	-1.45902860559818\\
-1	-2.45902860559818\\
0	-2.78382771499962\\
};
\addlegendentry{Right Extremal Point}

\end{axis}
\end{tikzpicture}%

%% file: images/centers.4.tex
\begin{tikzpicture}

\begin{axis}[%
width=7.083in,
height=5in,
at={(0.625in,0.625in)},
scale only axis,
xmin=-100,
xmax=0,
x tick label style={rotate=90,anchor=east},
xticklabel={\round[0]{10100+\tick}},
xlabel style={font=\color{white!15!black}},
xlabel={Iteration},
ymin=-10,
ymax=4,
ylabel style={font=\color{white!15!black}},
ylabel={Offset},
axis background/.style={fill=white},
title style={font=\bfseries},
axis x line*=bottom,
axis y line*=left,
legend style={at={(0.03,0.97)}, anchor=north west, legend cell align=left, align=left, draw=white!15!black}
]
\addplot[const plot, color=blue, line width=2pt] table[row sep=crcr] {%
-99	-2.05897502221254\\
-94	-2.05089076629933\\
-90	-2.05918757586204\\
-88	-2.05088714741252\\
-86	-2.05089076629933\\
-84	-2.05918757586204\\
-83	-2.06744156624362\\
-80	-2.05904182378589\\
-79	-2.05074139533639\\
-75	-2.05089076629933\\
-74	-2.05929050875706\\
-65	-2.06744156624362\\
-61	-2.05918757586204\\
-60	-2.05089076629933\\
-58	-2.04261054352514\\
-57	-2.03428818347419\\
-54	-2.02599833897169\\
-53	-2.0177700598955\\
-49	-2.00947462612109\\
-48	-2.00127231970933\\
-45	-2.00956852329179\\
-44	-2.0177700598955\\
-40	-2.02605913458996\\
-39	-2.03428818347419\\
-36	-2.02599833897169\\
-35	-2.0177700598955\\
-34	-2.02605990439801\\
-33	-2.02605913458996\\
-32	-2.0177700598955\\
-31	-2.00947385631304\\
-30	-2.00127231970933\\
-19	-2.00956852329179\\
-18	-2.00947462612109\\
-17	-2.00127231970933\\
-15	-2.00947462612109\\
-14	-2.0177700598955\\
-9	-2.02605990439801\\
-8	-2.03440247012428\\
-2	-2.03428818347419\\
0	-2.05897502221254\\
};
\addlegendentry{Core-Center}

\addplot[const plot, color=red, dashdotted, line width=2pt] table[row sep=crcr] {%
-99	-2.53244111070588\\
-98	-3.53244111070588\\
-97	-4.53244111070588\\
-96	-3.53244111070588\\
-95	-2.53244111070588\\
-94	-2.52019209619817\\
-93	-2.53244111070588\\
-92	-2.52019209619817\\
-91	-2.53244111070588\\
-90	-3.53244111070588\\
-89	-2.53244111070588\\
-88	-2.52019209619817\\
-87	-3.52019209619817\\
-86	-2.52019209619817\\
-85	-2.53244111070588\\
-84	-3.53244111070588\\
-83	-2.53287175823743\\
-82	-2.55021625529785\\
-81	-2.53287175823743\\
-80	-2.53244111070588\\
-79	-2.52019209619817\\
-78	-3.52019209619817\\
-77	-4.52019209619817\\
-76	-5.52019209619817\\
-75	-6.52019209619817\\
-74	-7.52019209619817\\
-73	-8.52019209619817\\
-72	-7.52019209619817\\
-71	-6.52019209619817\\
-70	-5.52019209619817\\
-69	-4.52019209619817\\
-68	-3.52019209619817\\
-67	-4.52019209619817\\
-66	-3.52019209619817\\
-65	-2.53287175823743\\
-64	-2.55021625529785\\
-63	-2.53287175823743\\
-62	-2.55021625529785\\
-61	-2.53244111070588\\
-60	-2.52019209619817\\
-59	-2.53244111070588\\
-58	-2.51778762083559\\
-57	-2.51055294226143\\
-56	-3.51055294226143\\
-55	-2.51055294226143\\
-54	-2.4970444380594\\
-53	-2.49695283090132\\
-52	-2.4970444380594\\
-51	-3.4970444380594\\
-50	-2.4970444380594\\
-49	-2.48420105721435\\
-48	-2.47302729390142\\
-47	-2.48420105721435\\
-46	-2.47302729390142\\
-45	-3.47302729390142\\
-44	-2.49695283090132\\
-43	-2.4970444380594\\
-42	-3.4970444380594\\
-41	-4.4970444380594\\
-40	-3.4970444380594\\
-39	-2.51055294226143\\
-38	-2.51778762083559\\
-37	-2.51055294226143\\
-36	-2.4970444380594\\
-35	-2.49695283090132\\
-34	-3.49695283090132\\
-33	-2.4970444380594\\
-32	-2.49695283090132\\
-31	-2.48420105721435\\
-30	-2.47302729390142\\
-29	-3.47302729390142\\
-28	-2.47302729390142\\
-27	-2.48420105721435\\
-26	-2.47302729390142\\
-25	-3.47302729390142\\
-24	-2.47302729390142\\
-23	-3.47302729390142\\
-22	-4.47302729390142\\
-21	-3.47302729390142\\
-20	-2.47302729390142\\
-19	-3.47302729390142\\
-18	-2.48420105721435\\
-17	-2.47302729390142\\
-16	-2.48420105721435\\
-15	-3.48420105721435\\
-14	-2.49695283090132\\
-13	-2.4970444380594\\
-12	-2.49695283090132\\
-11	-3.49695283090132\\
-10	-4.49695283090132\\
-9	-5.49695283090132\\
-8	-6.49695283090132\\
-7	-5.49695283090132\\
-6	-4.49695283090132\\
-5	-5.49695283090132\\
-4	-4.49695283090132\\
-3	-3.49695283090132\\
-2	-2.51055294226143\\
-1	-2.51778762083559\\
0	-3.53244111070588\\
};
\addlegendentry{Left Extremal Point}

\addplot[const plot, color=purple, dashed, line width=2pt] table[row sep=crcr] {%
-99	1.44183435747215\\
-98	2.44183435747215\\
-97	1.44183435747215\\
-96	0.441834357472146\\
-95	-0.558165642527854\\
-94	-1.53244111070588\\
-93	-1.52019209619817\\
-92	-1.53244111070588\\
-91	-1.52019209619817\\
-90	-1.53287175823743\\
-89	-0.532871758237434\\
-88	0.467128241762566\\
-87	-0.532871758237434\\
-86	-1.53244111070588\\
-85	-1.52019209619817\\
-84	-1.53287175823743\\
-83	-1.55021625529785\\
-82	-1.53287175823743\\
-81	-1.55021625529785\\
-80	-0.550216255297855\\
-79	0.449783744702145\\
-78	1.44978374470215\\
-77	0.449783744702145\\
-76	-0.550216255297855\\
-75	-1.53244111070588\\
-74	-1.53287175823743\\
-73	-0.532871758237434\\
-72	0.467128241762566\\
-71	-0.532871758237434\\
-70	-1.53287175823743\\
-69	-0.532871758237434\\
-68	-1.53287175823743\\
-67	-0.532871758237434\\
-66	-1.53287175823743\\
-65	-1.55021625529785\\
-64	-1.53287175823743\\
-63	-1.55021625529785\\
-62	-1.53287175823743\\
-61	-0.532871758237434\\
-60	-1.53244111070588\\
-59	-1.52019209619817\\
-58	-0.520192096198173\\
-57	-1.51778762083559\\
-56	-0.517787620835591\\
-55	-1.51778762083559\\
-54	-0.517787620835591\\
-53	-1.4970444380594\\
-52	-1.49695283090132\\
-51	-0.496952830901321\\
-50	-1.49695283090132\\
-49	-0.496952830901321\\
-48	-1.48420105721435\\
-47	-1.47302729390142\\
-46	-1.48420105721435\\
-45	-1.49695283090132\\
-44	-1.4970444380594\\
-43	-1.49695283090132\\
-42	-0.496952830901321\\
-41	-1.49695283090132\\
-40	-1.51055294226143\\
-39	-1.51778762083559\\
-38	-1.51055294226143\\
-37	-1.51778762083559\\
-36	-0.517787620835591\\
-35	-1.4970444380594\\
-34	-1.51055294226143\\
-33	-0.510552942261427\\
-32	-1.4970444380594\\
-31	-0.497044438059405\\
-30	-1.48420105721435\\
-29	-0.484201057214349\\
-28	-1.48420105721435\\
-27	-1.47302729390142\\
-26	-1.48420105721435\\
-25	-0.484201057214349\\
-24	0.515798942785651\\
-23	-0.484201057214349\\
-22	0.515798942785651\\
-21	-0.484201057214349\\
-20	-1.48420105721435\\
-19	-1.49695283090132\\
-18	-0.496952830901321\\
-17	-1.48420105721435\\
-16	-1.47302729390142\\
-15	-1.49695283090132\\
-14	-1.4970444380594\\
-13	-1.49695283090132\\
-12	-1.4970444380594\\
-11	-0.497044438059405\\
-10	-1.4970444380594\\
-9	-1.51055294226143\\
-8	-1.51778762083559\\
-7	-0.517787620835591\\
-6	-1.51778762083559\\
-5	-0.517787620835591\\
-4	-1.51778762083559\\
-3	-0.517787620835591\\
-2	-1.51778762083559\\
-1	-1.51055294226143\\
0	0.441834357472146\\
};
\addlegendentry{Right Extremal Point}

\end{axis}
\end{tikzpicture}%

%% file: images/FreddyNewModel.tex
\NewDocumentCommand{\bissectrice}{%
    O{}     
    mmm     
    m       
    O{1}O{1}
    }{%
    \path[name path=Bis#2#3#4] let
        \p1 = ($(#2) - (#3)$),
        \p2 = ($(#4) - (#3)$),
        \n1 = {veclen(\x1,\y1)/2} ,
        \n2 = {veclen(\x2,\y2)/2} ,
        \n3 = {max(\n1,\n2)},
        \p1 = ($(#3)!\n3!(#2)$),
        \p2 = ($(#3)!\n3!(#4)$),
        \p3 = ($(\p1) + (\p2) - (#3)$)
    in
        (#3) -- (\p3) ;

    \path[name path = foo] (#2)--(#4) ;

    \path[name intersections={of=foo and Bis#2#3#4, by=#5}] ;

    \path[#1] ($(#3)!#6!(#5)$) -- ($(#5)!#7!(#3)$) ;
    }

	\begin{tikzpicture}
		\coordinate (A) at (0, 0);
		\coordinate (B) at (-2, 2);
		\coordinate (C) at (-1.5, 4);
		\coordinate (D) at (1, 6);
		\coordinate (E) at (3, 5);
		\coordinate (F) at (4, 2);
		\coordinate (EP) at (5, 5);
		\coordinate (IP) at (5, 4);

		\bissectrice[draw, black]  {B}{A}{F}{R1}[0.6]
		\bissectrice[draw, black]  {B}{A}{F}{R2}[-0.6]
		
		\begin{scope}[spy using outlines={circle, magnification=2.2, connect spies}]
			\draw (A) 
				-- (B)
				-- (C)
				-- (D)
				-- (E)
				-- (F) 
				-- cycle;
				
			\foreach \i in { A, B, C, D, E, F} 
			{
				\node[very thin, circle, fill, inner sep = 1pt] (p_node_\i) at ({\i}) {};
			}
			
			\node[below right] (p_label) at (A) {P};
			
			\draw[-latex] (A) -- ($(A)!0.3!(R1)$);
			\draw[-latex] (A) -- ($(A)!-0.3!(R1)$);
		
			\draw[-latex] (EP) -- ($(E) + (0.2, 0)$);
			\draw[-latex] (EP) -- ($(D) + (0.2, -0.03)$);
			\node[right] (EP_label) at (EP) {Extremal points};
		
			\spy [blue, size=3.7cm] on (A)
				in node[fill=white] (SA) at (6, 1.5);
		\end{scope}

		\foreach \x/\y in {-1/2, -1.5/3, 0/1.7, 0.1/3.2, 0.2/5, 2/3, 2.4/4.8, 0.2/0.2, 1.2/4, 1.4/1, 0.8/2.4, 3.2/1.9}
		{
				\node[very thin, circle, fill, inner sep = 1pt] () at ({\x, \y}) {};
		}
		\draw[-latex] (IP) -- (1.4, 4);
		\draw[-latex] (IP) -- (1.6, 1.15);
		\draw[-latex] (IP) -- (2.2, 3.05);
		\node[right] (IP_label) at (IP) {Internal points};

		\coordinate (InMagnifyingGlass) at ($(SA)!0.3!($(SA) + (R1) - (A)$)$);
		
		\draw[dashed] (SA) circle (1.32cm);
		\draw[latex-] ($(SA) + (180:1.32)$) -- ($(SA) + (180:0.1)$);
		\node[above] (unit_label) at ($(SA) + (180:0.66)$) {$1$};
		
		\draw[latex-latex] (InMagnifyingGlass) arc (80:45:1);
		\draw[latex-latex] (InMagnifyingGlass) arc (80:115:1);
		
		\node[above right] at ($(SA) + (63:0.5)$) {\tiny $\nicefrac{\alpha}{2}$};
		\node[above left] at ($(SA) + (92:0.5)$) {\tiny $\nicefrac{\alpha}{2}$};
		
		\node[above right] at ($(SA) + (80:1)$) {\small $(1-\varepsilon)$};
		\node[below right] at ($(SA) + (-100:1)$) {\small $\varepsilon$};
	\end{tikzpicture}

%% file: images/FreddyConvexHulls.tex
	\begin{figure*}[h]
		\centering
		\begin{subfigure}[b]{0.49\textwidth}
			\resizebox{\linewidth}{!}{\includegraphics{{../pdfimages/ModelConvexHull.00000}.pdf}}
			\caption[]
			{{\small At the beginning}}
			\label{sub:SystemConvexHullAt000}
		\end{subfigure}
		\hfill
		\begin{subfigure}[b]{0.49\textwidth}
			\resizebox{\linewidth}{!}{\includegraphics{{../pdfimages/ModelConvexHull.00100}.pdf}}
			\caption[]
			{{\small After $100$ iterations}}
			\label{sub:SystemConvexHullAt100}
		\end{subfigure}
		\qquad
		\begin{subfigure}[b]{0.49\textwidth}
			\resizebox{\linewidth}{!}{\includegraphics{{../pdfimages/ModelConvexHull.00200}.pdf}}
			\caption[]
			{{\small After $200$ iterations}}
			\label{sub:SystemConvexHullAt200}
		\end{subfigure}
		\hfill
		\begin{subfigure}[b]{0.49\textwidth}
			\resizebox{\linewidth}{!}{\includegraphics{{../pdfimages/ModelConvexHull.00400}.pdf}}
			\caption[]
			{{\small After $400$ iterations}}
			\label{sub:SystemConvexHullAt400}
		\end{subfigure}
		\caption[ ]
		{\small Typical evolution of the system ($N = 400, \varepsilon = 0.1$) from (\subref{sub:SystemConvexHullAt000}) beginning till (\subref{sub:SystemConvexHullAt400}) $400^{th}$ iteration. (Convex Hull is depicted for convenience)} 
		\label{2DFreddyModelEvolution}
	\end{figure*}

%% file: Notes.bbl
\begin{thebibliography}{}

\bibitem[\protect\citeauthoryear{Ando, Oasa, Suzuki, and Yamashita}{Ando
  et~al.}{1999}]{ando1999distributed}
Ando, H., Y.~Oasa, I.~Suzuki, and M.~Yamashita (1999).
\newblock Distributed memoryless point convergence algorithm for mobile robots
  with limited visibility.
\newblock {\em IEEE Transactions on Robotics and Automation\/}~{\em 15\/}(5),
  818--828.

\bibitem[\protect\citeauthoryear{Barel, Manor, and Bruckstein}{Barel
  et~al.}{2016}]{barel2016}
Barel, A., R.~Manor, and A.~M. Bruckstein (2016, July).
\newblock {COME TOGETHER: Multi-Agent Geometric Consensus (Gathering,
  Rendezvous, Clustering, Aggregation)}.
\newblock Technical Report CIS-2016-03, {CS Department Technion IIT}.

\bibitem[\protect\citeauthoryear{Camazine, Deneubourg, Franks, Sneyd,
  Theraulaz, and Bonabeau}{Camazine et~al.}{2001}]{citeulike:606465}
Camazine, S., J.-L. Deneubourg, N.~R. Franks, J.~Sneyd, G.~Theraulaz, and
  E.~Bonabeau (2001, August).
\newblock {\em {Self-Organization} in Biological Systems}.
\newblock Princeton University Press.

\bibitem[\protect\citeauthoryear{Cartwright and Harary}{Cartwright and
  Harary}{1956}]{cartwright1956structural}
Cartwright, D. and F.~Harary (1956).
\newblock Structural balance: a generalization of heider's theory.
\newblock {\em Psychological Review\/}~{\em 63\/}(5), 277.

\bibitem[\protect\citeauthoryear{Chen, Jamieson, Balakrishnan, and Morris}{Chen
  et~al.}{2002}]{chen2002span}
Chen, B., K.~Jamieson, H.~Balakrishnan, and R.~Morris (2002).
\newblock Span: An energy-efficient coordination algorithm for topology
  maintenance in ad hoc wireless networks.
\newblock {\em Wireless Networks\/}~{\em 8\/}(5), 481--494.

\bibitem[\protect\citeauthoryear{Chong and Kumar}{Chong and
  Kumar}{2003}]{chong2003sensor}
Chong, C.-Y. and S.~P. Kumar (2003).
\newblock Sensor networks: evolution, opportunities, and challenges.
\newblock {\em Proceedings of the IEEE\/}~{\em 91\/}(8), 1247--1256.

\bibitem[\protect\citeauthoryear{Couzin and Krause}{Couzin and
  Krause}{2003}]{couzin2003self}
Couzin, I. and J.~Krause (2003).
\newblock Self-organization and collective behavior in invertebrates.
\newblock {\em Advances in the Study of Behavior\/}~{\em 32}, 1--75.

\bibitem[\protect\citeauthoryear{DeGroot}{DeGroot}{1974}]{degroot1974reaching}
DeGroot, M.~H. (1974).
\newblock Reaching a consensus.
\newblock {\em Journal of the American Statistical Association\/}~{\em
  69\/}(345), 118--121.

\bibitem[\protect\citeauthoryear{Dudek, Jenkin, Milios, and Wilkes}{Dudek
  et~al.}{1993}]{dudek1993taxonomy}
Dudek, G., M.~Jenkin, E.~Milios, and D.~Wilkes (1993).
\newblock A taxonomy for swarm robots.
\newblock In {\em Intelligent Robots and Systems' 93, IROS'93. Proceedings of
  the 1993 IEEE/RSJ International Conference on}, Volume~1, pp.\  441--447.
  IEEE.

\bibitem[\protect\citeauthoryear{Festinger}{Festinger}{1954}]{festinger1954theory}
Festinger, L. (1954).
\newblock A theory of social comparison processes.
\newblock {\em Human relations\/}~{\em 7\/}(2), 117--140.

\bibitem[\protect\citeauthoryear{Festinger}{Festinger}{1962}]{festinger1962theory}
Festinger, L. (1962).
\newblock {\em A theory of cognitive dissonance}, Volume~2.
\newblock Stanford University Press.

\bibitem[\protect\citeauthoryear{Flocchini, Prencipe, and Santoro}{Flocchini
  et~al.}{2012}]{flocchini2012distributed}
Flocchini, P., G.~Prencipe, and N.~Santoro (2012).
\newblock Distributed computing by oblivious mobile robots.
\newblock {\em Synthesis lectures on distributed computing theory\/}~{\em
  3\/}(2), 1--185.

\bibitem[\protect\citeauthoryear{Friedkin and Johnsen}{Friedkin and
  Johnsen}{1990}]{friedkin1990social}
Friedkin, N.~E. and E.~C. Johnsen (1990).
\newblock Social influence and opinions.
\newblock {\em Journal of Mathematical Sociology\/}~{\em 15\/}(3-4), 193--206.

\bibitem[\protect\citeauthoryear{Halpern and Moses}{Halpern and
  Moses}{1990}]{halpern1990knowledge}
Halpern, J.~Y. and Y.~Moses (1990).
\newblock Knowledge and common knowledge in a distributed environment.
\newblock {\em Journal of the ACM (JACM)\/}~{\em 37\/}(3), 549--587.

\bibitem[\protect\citeauthoryear{Hegselmann and Krause}{Hegselmann and
  Krause}{2002}]{hegselmann2002opinion}
Hegselmann, R. and U.~Krause (2002).
\newblock Opinion dynamics and bounded confidence models, analysis, and
  simulation.
\newblock {\em Journal of Artificial Societies and Social Simulation\/}~{\em
  5\/}(3).

\bibitem[\protect\citeauthoryear{Hilton and Pedersen}{Hilton and
  Pedersen}{1991}]{hilton1991catalan}
Hilton, P. and J.~Pedersen (1991).
\newblock Catalan numbers, their generalization, and their uses.
\newblock {\em The Mathematical Intelligencer\/}~{\em 13\/}(2), 64--75.

\bibitem[\protect\citeauthoryear{Jadbabaie, Lin, and Morse}{Jadbabaie
  et~al.}{2003}]{jadbabaie2003coordination}
Jadbabaie, A., J.~Lin, and A.~S. Morse (2003).
\newblock Coordination of groups of mobile autonomous agents using nearest
  neighbor rules.
\newblock {\em IEEE Transactions on Automatic Control\/}~{\em 48\/}(6),
  988--1001.

\bibitem[\protect\citeauthoryear{Krishna, Vaidya, Chatterjee, and
  Pradhan}{Krishna et~al.}{1997}]{krishna1997cluster}
Krishna, P., N.~H. Vaidya, M.~Chatterjee, and D.~K. Pradhan (1997).
\newblock A cluster-based approach for routing in dynamic networks.
\newblock {\em ACM SIGCOMM Computer Communication Review\/}~{\em 27\/}(2),
  49--64.

\bibitem[\protect\citeauthoryear{Lorenz}{Lorenz}{2017}]{lorenz2017modeling}
Lorenz, J. (2017).
\newblock Modeling the evolution of ideological landscapes through opinion
  dynamics.
\newblock In {\em Advances in Social Simulation 2015}, pp.\  255--266.
  Springer.

\bibitem[\protect\citeauthoryear{Okubo}{Okubo}{1986}]{citeulike:6170389}
Okubo, A. (1986).
\newblock Dynamical aspects of animal grouping: Swarms, schools, flocks, and
  herds.
\newblock {\em Advances in Biophysics\/}~{\em 22}, 1--94.

\bibitem[\protect\citeauthoryear{Olfati-Saber and Murray}{Olfati-Saber and
  Murray}{2004}]{olfati2004consensus}
Olfati-Saber, R. and R.~M. Murray (2004).
\newblock Consensus problems in networks of agents with switching topology and
  time-delays.
\newblock {\em IEEE Transactions on automatic control\/}~{\em 49\/}(9),
  1520--1533.

\bibitem[\protect\citeauthoryear{Reynolds}{Reynolds}{1987}]{reynolds1987flocks}
Reynolds, C.~W. (1987).
\newblock Flocks, herds and schools: A distributed behavioral model.
\newblock {\em ACM SIGGRAPH computer graphics\/}~{\em 21\/}(4), 25--34.

\bibitem[\protect\citeauthoryear{{\c{S}}ahin}{{\c{S}}ahin}{2004}]{csahin2004swarm}
{\c{S}}ahin, E. (2004).
\newblock Swarm robotics: From sources of inspiration to domains of
  application.
\newblock In {\em International Workshop on Swarm Robotics}, pp.\  10--20.
  Springer.

\bibitem[\protect\citeauthoryear{Shoham and Tennenholtz}{Shoham and
  Tennenholtz}{1995}]{shoham1995social}
Shoham, Y. and M.~Tennenholtz (1995).
\newblock On social laws for artificial agent societies: off-line design.
\newblock {\em Artificial intelligence\/}~{\em 73\/}(1-2), 231--252.

\bibitem[\protect\citeauthoryear{Stanley}{Stanley}{2015}]{stanley2015catalan}
Stanley, R.~P. (2015).
\newblock {\em Catalan numbers}.
\newblock Cambridge University Press.

\bibitem[\protect\citeauthoryear{Sumpter}{Sumpter}{2006}]{sumpter2006principles}
Sumpter, D.~J. (2006).
\newblock The principles of collective animal behaviour.
\newblock {\em Philosophical Transactions of the Royal Society of London B:
  Biological Sciences\/}~{\em 361\/}(1465), 5--22.

\bibitem[\protect\citeauthoryear{Vicsek, Czir{\'o}k, Ben-Jacob, Cohen, and
  Shochet}{Vicsek et~al.}{1995}]{vicsek1995novel}
Vicsek, T., A.~Czir{\'o}k, E.~Ben-Jacob, I.~Cohen, and O.~Shochet (1995).
\newblock Novel type of phase transition in a system of self-driven particles.
\newblock {\em Physical Review Letters\/}~{\em 75\/}(6), 1226.

\end{thebibliography}
